\documentclass[3p]{elsarticle}

\usepackage[color=yellow]{todonotes}
\usepackage{float}
\usepackage[title]{appendix}
\usepackage{graphicx}
\usepackage{caption,subcaption}
\usepackage{siunitx}
\usepackage{multirow}
\usepackage{algorithm}
\usepackage{algorithmic,multicol}
\usepackage{bm}
\usepackage[parfill]{parskip}
\usepackage[utf8]{inputenc}
\usepackage{url}
\usepackage{amsmath,amssymb,amsfonts,amsthm}
\usepackage{color}
\usepackage{bm}
\usepackage{epstopdf}
\usepackage{array}
\usepackage[hidelinks]{hyperref}
\usepackage{cleveref}
\usepackage{lineno}
\usepackage{bbm}
\usepackage{comment}
\usepackage{xfrac}
\usepackage{varwidth}
\usepackage[nice]{nicefrac}

\makeatletter
\def\maketag@@@#1{\hbox{\m@th\normalfont\normalsize#1}}
\makeatother

\ifpdf
\DeclareGraphicsExtensions{-eps-converted-to.pdf,.pdf,.png,.jpg}
\else
\DeclareGraphicsExtensions{-eps-converted-to.pdf}
\fi

\newtheorem{proposition}{Proposition}

\makeatletter
\newsavebox\myboxA
\newsavebox\myboxB
\newlength\mylenA

\newcommand*\xoverline[2][0.75]{%
	\sbox{\myboxA}{$\m@th#2$}%
	\setbox\myboxB\null
	\ht\myboxB=\ht\myboxA%
	\dp\myboxB=\dp\myboxA%
	\wd\myboxB=#1\wd\myboxA
	\sbox\myboxB{$\m@th\overline{\copy\myboxB}$}
	\setlength\mylenA{\the\wd\myboxA}
	\addtolength\mylenA{-\the\wd\myboxB}%
	\ifdim\wd\myboxB<\wd\myboxA%
	\rlap{\hskip 0.5\mylenA\usebox\myboxB}{\usebox\myboxA}%
	\else
	\hskip -0.5\mylenA\rlap{\usebox\myboxA}{\hskip 0.5\mylenA\usebox\myboxB}%
	\fi}
\makeatother

\usepackage{tikz}
\usetikzlibrary{chains, positioning, arrows.meta, bending, shapes.arrows}
\usetikzlibrary{positioning,shapes,arrows,calc,shapes.geometric,backgrounds,fit}
\usepackage{pifont}
\usetikzlibrary{positioning,shapes,arrows,calc,shapes.geometric,backgrounds,fit}

\tikzstyle{block} = [rectangle,draw,minimum width=2em,align=center,rounded corners, minimum height=2em,scale=1.0]
\tikzstyle{blockleft} = [rectangle,draw,minimum width=2em,align=left,rounded corners, minimum height=2em,scale=1.0]
\tikzstyle{bigblock} = [rectangle,draw,minimum width=8em,align=center,rounded corners, minimum height=4em,scale=1.0]
\tikzstyle{connect} = [draw,-latex']
\tikzstyle{decision} = [diamond, draw, 
text width=4.5em, text badly centered, node distance=3cm, inner sep=0pt]
\tikzstyle{line} = [draw, -latex']
\tikzstyle{cloud} = [draw, ellipse,fill=red!20, node distance=3cm,
minimum height=2em]
\tikzstyle{linenoarrow}=[draw]

\makeatletter
\let\save@mathaccent\mathaccent
\newcommand*\if@single[3]{%
	\setbox0\hbox{${\mathaccent"0362{#1}}^H$}%
	\setbox2\hbox{${\mathaccent"0362{\kern0pt#1}}^H$}%
	\ifdim\ht0=\ht2 #3\else #2\fi
}
\newcommand*\rel@kern[1]{\kern#1\dimexpr\macc@kerna}
\newcommand*\widebar[1]{\@ifnextchar^{{\wide@bar{#1}{0}}}{\wide@bar{#1}{1}}}
\newcommand*\wide@bar[2]{\if@single{#1}{\wide@bar@{#1}{#2}{1}}{\wide@bar@{#1}{#2}{2}}}
\newcommand*\wide@bar@[3]{%
	\begingroup
	\def\mathaccent##1##2{%
		\let\mathaccent\save@mathaccent
		\if#32 \let\macc@nucleus\first@char \fi
		\setbox\z@\hbox{$\macc@style{\macc@nucleus}_{}$}%
		\setbox\tw@\hbox{$\macc@style{\macc@nucleus}{}_{}$}%
		\dimen@\wd\tw@
		\advance\dimen@-\wd\z@
		\divide\dimen@ 3
		\@tempdima\wd\tw@
		\advance\@tempdima-\scriptspace
		\divide\@tempdima 10
		\advance\dimen@-\@tempdima
		\ifdim\dimen@>\z@ \dimen@0pt\fi
		\rel@kern{0.6}\kern-\dimen@
		\if#31
		\overline{\rel@kern{-0.6}\kern\dimen@\macc@nucleus\rel@kern{0.4}\kern\dimen@}%
		\advance\dimen@0.4\dimexpr\macc@kerna
		\let\final@kern#2%
		\ifdim\dimen@<\z@ \let\final@kern1\fi
		\if\final@kern1 \kern-\dimen@\fi
		\else
		\overline{\rel@kern{-0.6}\kern\dimen@#1}%
		\fi
	}%
	\macc@depth\@ne
	\let\math@bgroup\@empty \let\math@egroup\macc@set@skewchar
	\mathsurround\z@ \frozen@everymath{\mathgroup\macc@group\relax}%
	\macc@set@skewchar\relax
	\let\mathaccentV\macc@nested@a
	\if#31
	\macc@nested@a\relax111{#1}%
	\else
	\def\gobble@till@marker##1\endmarker{}%
	\futurelet\first@char\gobble@till@marker#1\endmarker
	\ifcat\noexpand\first@char A\else
	\def\first@char{}%
	\fi
	\macc@nested@a\relax111{\first@char}%
	\fi
	\endgroup
}
\makeatother
\crefname{appendix}{}{}
\Crefname{figure}{Figure}{Figures}
\Crefname{section}{Section}{Sections}
\Crefname{equation}{Equation}{Equations}
\Crefname{table}{Table}{Tables}
\usepackage{stackengine}
\usepackage[nopar]{lipsum}
\newcommand\scalemath[2]{\scalebox{#1}{\mbox{\ensuremath{\displaystyle #2}}}}

\stackMath
\def\simiid{\mathrel{\stackon[4pt]{\sim}{\scalemath{0.7}{i.i.d.}}}}
\begin{document}
\begin{frontmatter}
	\title{Multivariate error modeling and uncertainty quantification using importance (re-)weighting for Monte Carlo simulations in particle transport\\[0.5em] \normalsize   \vspace{0em}}

	\author[adress1,adress2,adress3]{Pia Stammer}
	\author[adress2,adress4]{Lucas Burigo}
	\author[adress2,adress3,adress4,adress5]{Oliver Jäkel}
	\author[adress1,adress3]{Martin Frank}
	\author[adress2,adress4]{Niklas Wahl}
	
	\address[adress1]{Karlsruhe Institute of Technology, Steinbuch Centre for Computing,
		Karlsruhe, Germany}
	\address[adress2]{German Cancer Research Center - DKFZ, Department of Medical Physics in
		Radiation Oncology, Heidelberg, Germany}
	\address[adress3]{HIDSS4Health - Helmholtz Information and Data Science School for Health,
		Karlsruhe/Heidelberg, Germany}
	\address[adress4]{Heidelberg Institute for Radiation Oncology (HIRO), Heidelberg, Germany}
	\address[adress5]{Heidelberg Ion Beam Therapy Center — HIT, Department of Medical Physics in Radiation Oncology, Heidelberg, Germany}
\journal{}
	
	\begin{abstract} 
		
		Fast and accurate predictions of uncertainties in the computed dose are crucial for the determination of robust treatment plans in radiation therapy. This requires the solution of particle transport problems with uncertain parameters or initial conditions. Monte Carlo methods are often used to solve transport problems especially for applications which require high accuracy. In these cases, common non-intrusive solution strategies that involve repeated simulations of the problem at different points in the parameter space quickly become infeasible due to their long run-times. Intrusive methods however limit the usability in combination with proprietary simulation engines. In \cite{stammerEfficientUncertaintyQuantification2021}, we demonstrated the application of a new non-intrusive uncertainty quantification approach for Monte Carlo simulations in proton dose calculations with normally distributed errors on realistic patient data.  In this paper, we introduce a generalized formulation and focus on a more in-depth theoretical analysis of this method concerning bias, error and convergence of the estimates. The multivariate input model of the proposed approach further supports almost arbitrary error correlation models. We demonstrate how this framework can be used to model and efficiently quantify complex auto-correlated and time-dependent errors.
	\end{abstract}

	
\end{frontmatter}
	\section{Introduction}
	
	The linear Boltzmann equation has been used to model transport processes in a large variety of applications \cite[e.g.][]{perthameTransportEquationsBiology2007, spanierMonteCarloPrinciples1969, duderstadtTransportTheory1979, bedfordCalculationAbsorbedDose2019}. Computational methods in radiation therapy require a discretization of the seven-dimensional phase space. These discretizations can be grouped into deterministic and stochastic approaches. Deterministic approaches \cite{bedfordCalculationAbsorbedDose2019, frankApproximateModelsRadiative2007}, such as the spherical harmonics ($P_N$) \cite{davisonNeutronTransportTheory1957} and the discrete ordinates ($S_N$) \cite{vilhenaParticularSolutionSN1995, giffordComparisonFiniteelementMultigroup2006} method for the angular discretization, are computationally efficient. However, stochastic Monte Carlo methods are increasingly preferred for radiation therapy due to their accuracy and flexibility \cite{wengVectorizedMonteCarlo2003, jabbariReviewFastMonte2011}. They also have the advantage of a known error estimate, an intuitive physical interpretation and being well-suited for parallel implementation. Under the assumption of exact geometry, this theoretically enables an arbitrarily exact approximation at the cost of longer run-times. However, in radiation therapy, human error as well as lack of information can cause deviations of the treatment reality from the model. Therefore, in addition to the physical distribution of phase space parameters, uncertainties also have a significant impact on the outcome of the simulations \cite{lomaxIntensityModulatedProton2008, kraanDoseUncertaintiesIMPT2013}. 
	
	When using the Monte Carlo method to solve the transport problem, quantifying the impact of these uncertainties can be challenging, as many UQ methods, such as (quasi-) Monte Carlo \cite{caflischMonteCarloQuasiMonte1998} or stochastic collocation \cite{xiuHighOrderCollocationMethods2005}, require numerous repeated runs of the solver at different points in the uncertain space. Further, the incorporation and sampling of multivariate error correlation models, which are essential for a realistic representation of certain error types, can be extremely tedious in scenario-based methods. Most other approaches either cannot provide the required accuracy \cite[e.g.][]{fredrikssonScenariobasedGeneralizationRadiation2015, casiraghiAdvantagesLimitationsWorst2013}, are only aimed at deterministic solution methods \cite[e.g.][]{bangertAnalyticalProbabilisticModeling2013, perkoFastAccurateSensitivity2016} or are intrusive \cite[e.g.][]{poetteGPCintrusiveMonteCarloScheme2018}, which is not realizable with the often proprietary softwares used for Monte Carlo simulations.
	
	We introduce a non-intrusive method, that generates estimates of the expected value and variance of radiation therapy doses from a single MC simulation run. The approach is based on the concept of importance sampling \cite{kahnRandomSamplingMonte1950, hastingsMonteCarloSampling1970} and mimics the computation of scenarios in the parameter space by (re-)weighting the results of a single realization. The computational overhead of a complex simulation is thus reduced to that of the (re-)weighting step, effectively making the uncertainty quantification problem a scoring problem independent from the underlying physical simulation. 
	
	Similar strategies have been previously employed, e.g. to estimate the average of samples from different distributions \cite{beckmanMonteCarloEstimation1987,tukeyConfiguralPolysampling1987}, for more efficient updates in optimization \cite{shirakawaSampleReuseImportance2018} and reinforcement learning \cite{peshkinChristianSheltonLearning1992, hachiyaAdaptiveImportanceSampling2009} or to artificially increase the sample size in costly simulations \cite{supanitskyEffectMultipleReusing2008}. However, to the best of our knowledge there is no previous research on computing a conditional variance by (re-)weighting a single original sample. Also, in the field of medical physics and in particular dose calculation, we are not aware of any previous use of importance (re-)weighting for the reuse of samples, neither for dose estimates (which are averages) nor for the variance with respect to input uncertainties. Note that a detailed application of the proposed approach to intensity modulated proton therapy (IMPT) treatment plans and several patient cases from the open CORT (common optimization for radiation therapy) data set \cite{craftSharedDataIntensity2014} is described in our previous publication \cite{stammerEfficientUncertaintyQuantification2021}.
	
	In the following, we first introduce the problem of particle transport with uncertain input in the context of radiation therapy (\cref{sec:radiationTransport}), then show how to use importance sampling to quantify uncertainties (\cref{sec:(Re-)weightingMCSamples}), discuss mathematical properties of this approach (\cref{sec:mathematicalProperties}) as well as different models for the input uncertainties (\cref{sec:modelingUncertainties}), and lastly show results for a water phantom as well as a prostate patient (\cref{sec:results}).

	\section{Solution of the deterministic linear Boltzmann equation using Monte Carlo}
	\label{sec:radiationTransport}
	
	We are interested in particle transport for an application in radiation therapy. The behavior of particles can be described by the Boltzmann transport equation. Here we focus on protons, which are especially sensitive to uncertainties due to the sharp peak in their depth-dose curve \cite{lomaxIntensityModulatedProton2008, liuRobustOptimizationIntensity2012}. 
	
	Let $\mathcal{D}_r$ be a bounded region containing the voxelized patient anatomy and $\Phi(\boldsymbol{r},\boldsymbol{\Omega},E)$ the proton fluence at location $\boldsymbol{r}\in\mathcal{D}_r\subset\mathbb{R}^3$, with unit velocity direction $\boldsymbol{\Omega}\in\mathbb{S}^2$ and energy $E\in\mathbb{R}_+$. The linear Boltzmann equation reads

	\begin{align}
	\label{BoltzmannEq}
	\Omega \cdot \nabla \Phi(\boldsymbol{r},\boldsymbol{\Omega},E) + \sigma_t(\boldsymbol{r},E) \Phi(\boldsymbol{r},\boldsymbol{\Omega},E) &= Q^{Sca}(\boldsymbol{r},\boldsymbol{\Omega},E) + S(\boldsymbol{r},\boldsymbol{\Omega},E) \;,
	\end{align}
	
    where $Q^{Sca}$ is the energy loss and scattering component and is given as 
	
	\begin{equation}
	Q^{Sca}(\boldsymbol{r}, \boldsymbol{\Omega},E) = \int_0^{\infty} \int_{\boldsymbol{\Omega'} \in S^2} \sigma_s(\boldsymbol{r},\boldsymbol{\Omega}\cdot \boldsymbol{\Omega'},E\rightarrow E') \Phi(\boldsymbol{r},\boldsymbol{\Omega'},E') d\boldsymbol{\Omega'} dE' \;.
	\end{equation}
	Note that the total and differential cross sections $\sigma_t$ and $\sigma_s$ are material and particle dependent and $S(\boldsymbol{r},\boldsymbol{\Omega},E)$ is the source function, which characterizes the spatial, angular, and energy distribution of the radiation source. The dose can then be determined from the fluence as 
	
	\begin{equation}
	D(\boldsymbol{r}) = \frac{1}{\rho(\boldsymbol{r})}  \int_0^{\infty} \int_{\boldsymbol{\Omega} \in S^2} L(\boldsymbol{r},E) \Phi(\boldsymbol{r},\boldsymbol{\Omega},E) d\boldsymbol{\Omega} dE \;.
	\end{equation}
	Here, $L(\boldsymbol{r},E)$ is the stopping power, i.e., the rate at which particles lose energy and $\rho(\boldsymbol{r})$ is the tissue density at position $\boldsymbol{r}$.\\
	
	\subsection{The Monte Carlo method}
	The Monte Carlo method is a numerical integration method, which is based on evaluating the integrand at random positions sampled from the probability distribution of the integration variable. For realizations $\boldsymbol{x_i}\,,i=1,...,N$ of the random variable $\boldsymbol{X}\in\mathbb{R}^d$ with probability density function $p:\mathbb{R}^d\rightarrow\mathbb{R}_+$, the estimator is defined as follows:
	
	\begin{equation}
	\label{eq:MC}
	I(f) = E_p[f(\boldsymbol{X})] = \int_{\mathbb{R}^d} f(\boldsymbol{X}) p(\boldsymbol{X}) d\boldsymbol{X} \approx \frac{1}{N} \sum_{i=1}^{N} f(\boldsymbol{x_i}):= I^N(f) \;.
	\end{equation}
 Note, that with the \textit{Strong Law of Large Numbers}, one can show   
	
	\begin{equation}
    P\left(\lim_{N\rightarrow \infty}  I^N(f)  \,= I(f)\right)=1 \;,
	\end{equation}
	i.e., the Monte Carlo estimator converges to the true value in probability. Further, the independence of the random variables $\boldsymbol{x_i}$ and the \textit{Central Limit Theorem} provide a formula for the variance of the estimator:
	
	\begin{equation}
    Var(I^N(f)) = \frac{v_f^2}{N}\;,
	\end{equation}
	which depends solely upon the variance $v_f^2$ of the integrand and the number of samples $N$. The estimator thus converges with $O(\frac{1}{\sqrt{N}})$ and its accuracy can be controlled through the number of computed realizations. 
	
	\subsection{Randomized quasi-Monte Carlo}
	\label{sec:RQMC}
	Due to the slow convergence and therefore often large sample sizes required for Monte Carlo estimates, numerous variations of the method exist which reduce the variance of the estimator.\\
	The quasi-Monte Carlo method replaces the randomly sampled points with a set of deterministic \textit{low discrepancy} points $P_N:=\{\boldsymbol{u_1},...,\boldsymbol{u_N}\}$, which cover the domain more uniformly \cite{caflischMonteCarloQuasiMonte1998}:
	\begin{equation}
	\label{eq:qMC}
	I_{qMC}^N(f) := \frac{1}{N} \sum_{i=1}^N f(\boldsymbol{u_i}) \;.
	\end{equation}
	This can for example be achieved using grid points or number theoretic sequences such as the Sobol sequence \cite{sobol1967distribution}, used later in this paper. These typically result in a faster convergence, however it is less straightforward to compute the error compared to the standard Monte Carlo method \cite{lEcuyer2018randomized}.\\
	Randomized quasi-Monte Carlo therefore reintroduces randomness to the method, while retaining the convergence properties of quasi-MC. The resulting point set is a low discrepancy sequence, in which each individual point is uniformly distributed \cite{lEcuyer2018randomized}. In the case of a grid rule, this can be achieved e.g. rotating the lattice or randomly drawing points within the grid cells from a uniform distribution. For number theoretic sequences, different random shuffling and scrambling\footnote{Shuffling refers to a randomization of the sample index, while scrambling refers to the randomization of the sample value.} strategies have been developed \cite{owen1995randomly, matousek1998L2discrepancy,laine2011stratified,kollig2002efficient}. We use a Sobol sequence with linear scrambling and a random digit shift according to \cite{matousek1998L2discrepancy}. For square-integrable functions, the resulting randomized quasi-Monte Carlo approach preserves the properties of Sobol sequences, ensuring a variance of $O \left( \frac{\log(N)^d}{N} \right)$ for $N$ $d$-dimensional sample points, while also generating an unbiased estimate and allowing for a statistical estimation of the integration error \cite{owen1995randomly, owenMCVariance1997}. 
	
	\subsection{The Boltzmann equation in integral form}
	In order to solve \cref{BoltzmannEq} using a variant of the Monte Carlo method, it needs to be given in its integral form. We first execute a change of variables, where $\boldsymbol{r}$ is parametrized with respect to a reference location $\boldsymbol{r_0}$
	
	\begin{align}
	\boldsymbol{r} = \boldsymbol{r_0}+s \boldsymbol{\Omega} \;.
	\end{align}
	
	For all combinations of $\boldsymbol{r'}$  and $\boldsymbol{\Omega'}$, such a reference location $\boldsymbol{r_0'}$ is determined, s.t. $\boldsymbol{r_0'}+s\boldsymbol{\Omega'}$:
	
	\begin{align}
	&Q_{Sca}(\boldsymbol{r}, \boldsymbol{\Omega},E) = \int_0^{\infty} \int_{\boldsymbol{\Omega'} \in S^2} \sigma_s(\boldsymbol{r_0'}+s\boldsymbol{\Omega'},\boldsymbol{\Omega}\cdot \boldsymbol{\Omega'},E\rightarrow E') \Phi(\boldsymbol{r_0'}+s\boldsymbol{\Omega'},\boldsymbol{\Omega'},E') d\boldsymbol{\Omega'} dE' \;, \\
	&\frac{d}{ds}\Phi(\boldsymbol{r_0}+s\boldsymbol{\Omega},\boldsymbol{\Omega},E) + \sigma_t(\boldsymbol{r_0}+s\boldsymbol{\Omega},E) \Phi(\boldsymbol{r_0}+s\boldsymbol{\Omega},\boldsymbol{\Omega},E) = Q^{Sca}(\boldsymbol{r_0}+s\boldsymbol{\Omega},\boldsymbol{\Omega},E) + S(\boldsymbol{r_0}+s\boldsymbol{\Omega},\boldsymbol{\Omega},E)\;.
	\end{align}
	
	For simplicity we now rename $\boldsymbol{r_0} + s \boldsymbol{\Omega} \rightarrow s$,
	
	\begin{equation}
	\frac{d}{ds}\Phi(s,\boldsymbol{\Omega},E) + \sigma_t(s,E) \Phi(s,\boldsymbol{\Omega},E) = Q^{Sca}(s,\boldsymbol{\Omega},E)+ S(s,\boldsymbol{\Omega},E) \;.
	\end{equation}
	
	Lastly we multiply with the integrating factor $\exp(-\int_0^{s} \sigma_t(s',E)ds')$ and integrate both sides to obtain

	\begin{align}
	\label{eq:integralLBE}
	\nonumber \Phi(s,\boldsymbol{\Omega},E) = & \;\Phi(\boldsymbol{r_0},\boldsymbol{\Omega},E)\exp\left(-\int_0^{s} \sigma_t(s',E)ds'\right) \\	&+ \int_0^{s} \left(Q^{Sca}(s'',\boldsymbol{\Omega},E) + S(s'',\boldsymbol{\Omega},E)\right)\, \exp\left(-\int_{s''}^{s} \sigma_t(s',E)ds'\right) ds'' \;.
	\end{align}
	
	Now the transport problem can be solved by sampling particles with phase space coordinates $(\boldsymbol{r},\boldsymbol{\Omega},E)$ from the source function $S(\boldsymbol{r},\boldsymbol{\Omega},E)$ and simulating their trajectories and random interactions according to the respective probability distributions. This corresponds to interpreting \cref{eq:integralLBE} in the sense of \cref{eq:MC} while choosing the probability density function $p$ to represent the respective particle source and physics. For the most accurate solution, this simulation is done event by event for each individual particle, always sampling the distance to the next collision, the type of collision, energy loss and scattering angle \cite{vassilievBoltzmannEquation2017}. However, in order to increase computation speed, different variations of this method exist, such as condensed history algorithms which only explicitly simulate so called catastrophic collision events, while accounting for soft collisions collectively after a given traveled distance. We will not elaborate this further here, the interested reader might take a look at \cite{vassilievBoltzmannEquation2017} or \cite{poetteGPCintrusiveMonteCarloScheme2018} for a more concise derivation of the probability measures, the resulting solution algorithm and different Monte Carlo variations.
	
	When considering non-intrusive UQ methods, the Monte Carlo dose calculation algorithm can be treated as a black box $BB(\boldsymbol{r},\boldsymbol{\Omega},E)$, returning the dose $D(\boldsymbol{r})$ at position $\boldsymbol{r}$ for the input vector $\boldsymbol{X}=(\boldsymbol{r},\boldsymbol{\Omega},E)\sim S_0$ on $\mathcal{D}_{S_0}$. 
	\begin{equation}
	D(\boldsymbol{r})=E_{S_0}[BB(\boldsymbol{r},\boldsymbol{\Omega},E)]= \int_{\mathcal{D}_{S_0}} BB(\boldsymbol{X}) S_0(\boldsymbol{X}) d\boldsymbol{X} \approx \frac{1}{H}\sum_{p=1}^H BB(\boldsymbol{x_p}) \;,
	\end{equation}
	where the realizations $\boldsymbol{x_1},...,\boldsymbol{x_H}$ are sampled from $S_0(\boldsymbol{X})$.
	
	Note, that the black box for fixed realizations of the input parameters still describes a stochastic process, due to subsequent sampling of the path and interactions of the particles. The input sampled from the source distribution merely fixes the initial properties of released particles. Each response $BB(\boldsymbol{r},\boldsymbol{\Omega},E)$ however also implicitly includes realizations of the probability distributions governing the interactions of these particles. Thus, the Monte Carlo method is applied to the complete random particle trajectory, we however omit this, since we are only interested in uncertainties affecting the phase space at a superficial level in the source distribution.
	
	\section{Introducing uncertainties}
	We now introduce a random error vector $\boldsymbol{Z}$ with density $p_{Z}$ to the problem \cref{BoltzmannEq}. The corresponding uncertain linear Boltzmann equation takes the following  form.
	
	\begin{align}
	\label{UncBoltzmannEq}
	\Omega \cdot \nabla \Phi(\boldsymbol{r},\boldsymbol{\Omega},E,\boldsymbol{Z}) + \sigma_t(\boldsymbol{r},E,\boldsymbol{Z}) \Phi(\boldsymbol{r},\boldsymbol{\Omega},E,\boldsymbol{Z}) &= Q^{Sca}(\boldsymbol{r},\boldsymbol{\Omega},E,\boldsymbol{Z}) + S(\boldsymbol{r},\boldsymbol{\Omega},E,\boldsymbol{Z}) \;,
	\end{align}
	
	We are interested in the expected value and variance of the dose estimate with respect to uncertainties, which occur during radiation treatments.
	
	\subsection{Uncertainties in radiation therapy}
	\label{sec:introducingUncertainties}
	Among the most important sources of uncertainties in proton therapy are such that manifest in the patient set-up and particle range \cite{parkStatisticalAssessmentProton2013, liuRobustOptimizationIntensity2012,perkoFastAccurateSensitivity2016,lomaxIntensityModulatedProton2008, lomaxIntensityModulatedProton2008a}. Set-up uncertainties directly affect the particle source $S_0$, by shifting the initial particle positions $\boldsymbol{r}$ by a random amount $\boldsymbol{Z_r}$. Range uncertainties are caused by a variety of factors concerning the patient density, ranging from CT conversion errors to imaging artifacts or changes in the patient geometry \cite{unkelbachAccountingRangeUncertainties2007,paganettiRangeUncertaintiesProton2012, lomaxIntensityModulatedProton2008,mcgowanTreatmentPlanningOptimisation2013}. Therefore, they manifest in the material dependent quantities, such as the scattering cross sections $\sigma_s$ and $\sigma_t$, as well as the density $\rho$ and the stopping power $L$.
	
	In the following, the method will be introduced for a general random error vector $\boldsymbol{Z}$ which affects the initial particle parameters $\boldsymbol{X}(\boldsymbol{Z})$ and the corresponding phase space distribution. We assume that the conditional distribution $S_{0,\boldsymbol{X}|\boldsymbol{Z}}(\boldsymbol{X}|\boldsymbol{Z}=\boldsymbol{z})$ of $\boldsymbol{X}$ for a fixed error realization is known. An application of the proposed method to set-up uncertainties will be discussed in \cref{sec:modelingUncertainties}. 
	
	\subsection{Solution of the uncertain problem using Monte Carlo}
	
	Most non-intrusive strategies involve computing the dose at different points in the parameter space. In case of the Monte Carlo method, these points are realizations randomly drawn from the probability distribution $p_{Z}$ of the uncertain variable $\boldsymbol{Z}$.
	
	Note that the introduction of an uncertain parameter adds a second stochastic process around the black box transport solver. On the one hand, the dose distribution depends on random input parameters following a distribution defined by the physical radiation source. On the other hand, this distribution now depends on the random error vector $\boldsymbol{Z}$. Thus, if we want to compute the expected value $\mathcal{D}$ of the dose with respect to both processes, we are looking for 
	
	\begin{equation}
	\label{eq:targetQuantityExpectedValue}
	\mathcal{D}(\boldsymbol{r})= E_{p_Z} \left[ E_{S_{0,\boldsymbol{X}|\boldsymbol{Z}}} \left [BB(\boldsymbol{X}(\boldsymbol{Z})) |\boldsymbol{Z}=\boldsymbol{z} \right] \right] = \int_{\mathcal{D}_{p_Z}}\int_{\mathcal{D}_{S_0}} BB(\boldsymbol{X}(\boldsymbol{Z})) \cdot S_{0,\boldsymbol{X}|\boldsymbol{Z}}(\boldsymbol{X}|\boldsymbol{Z}=\boldsymbol{z}) \, d\boldsymbol{X} \cdot p_Z(\boldsymbol{Z}) \,d\boldsymbol{Z}
	\end{equation}
	
	The Monte Carlo method can now be applied to both random processes. Starting with the error variable, for a fixed realization, the black box MC solver can be applied analogously to the original problem. Thus, the dose estimate $D_i(\boldsymbol{r})$ for realization $\boldsymbol{z_i}$ and the corresponding conditional distribution $S_i=S_{0,\boldsymbol{X}|\boldsymbol{Z}}(\boldsymbol{X}|\boldsymbol{Z}=\boldsymbol{z_i})$ is given by
	\begin{equation}
	\label{eq:doserealizationMC}
	D_i(\boldsymbol{r})=E_{S_i}[BB(\boldsymbol{X}(\boldsymbol{Z}))|\boldsymbol{Z}=\boldsymbol{z_i}] =  \int_{\mathcal{D}_{S_i}} BB(\boldsymbol{X}(\boldsymbol{z_i})) S_i(\boldsymbol{X}|\boldsymbol{Z}=\boldsymbol{z_i}) d\boldsymbol{X} \approx \frac{1}{H} \sum_{p=1}^H BB(\boldsymbol{x_p^i})
	\end{equation}
	with $\boldsymbol{x^i_1},...,\boldsymbol{x^i_H}$ sampled from $S_i$. The Monte Carlo estimate of the expected value with respect to both processes is then computed as the mean over $D_1(\boldsymbol{r}),...,D_N(\boldsymbol{r})$ for $\boldsymbol{z_1},...,\boldsymbol{z_N}$ sampled from $p_{Z}$
	\begin{align}
	\mathcal{D}(\boldsymbol{r})  \approx \frac{1}{N}\sum_{i=1}^N D_i(\boldsymbol{r}) 
	\approx \frac{1}{N}\sum_{i=1}^N \frac{1}{H} \sum_{p=1}^H BB(\boldsymbol{x_p^i})\;, 
	\end{align}
	and the variance as the sample variance, defined as
	
	\begin{equation}
	    S_N^2 = \frac{1}{N-1} \sum_{i=1}^N (\boldsymbol{x_i} - \bar{\boldsymbol{x}})^2 \, \text{ where } \bar{\boldsymbol{x}}=\frac{1}{N} \sum_{i=1}^N \boldsymbol{x_i}\,,
	\end{equation}
	
	for $\boldsymbol{x_i},\; i=1,...,N$ that are independent and identically distributed. This is an unbiased estimator of the variance of the random variable $\boldsymbol{X}$. Applied to the problem at hand, this results in the following formula for the dose variance:
	
	\begin{align}
	\nonumber Var_{p_{\boldsymbol{Z}}}(D(\boldsymbol{r},\boldsymbol{Z}))&\approx \frac{1}{N-1}\sum_{i=1}^N \left(D_i(\boldsymbol{r})-\mathcal{D}(\boldsymbol{r})\right)^2 \\
	&\approx \frac{1}{N-1}\sum_{i=1}^N \left(\frac{1}{H} \sum_{p=1}^H BB(\boldsymbol{x_p^i}) - \frac{1}{N}\sum_{j=1}^N \frac{1}{H} \sum_{p=1}^H BB(\boldsymbol{x_p^j})\right)^2 \;.
	\end{align}

	\subsection{Direct computation of the expected value}
	\label{sec:directExpDose}
	When not only the conditional or marginal distributions of $\boldsymbol{X}$ and $\boldsymbol{Z}$, but also their joint distribution $\mathcal{S}(\boldsymbol{X}(\boldsymbol{Z}))$ is known, it is possible to compute the expected dose directly 
	\begin{equation}
	\label{eq:expectedDoseJoint}
	E_{p_{\boldsymbol{Z}}} [D(\boldsymbol{r},\boldsymbol{Z})] = \mathbb{E}_{\mathcal{S}}[BB(\boldsymbol{X}(\boldsymbol{Z}))] \approx \frac{1}{H}\sum_{p=1}^H BB(\boldsymbol{x_p}) \;,
	\end{equation}
	with $\boldsymbol{x}_p, \; p=1,..,H$ sampled from $\mathcal{S}$.

	\section{(Re-)weighting Monte Carlo samples for uncertainty quantification}
	\label{sec:(Re-)weightingMCSamples}
	
	In the following section, we introduce the proposed (re-)weighting approach. We exploit the fact, that the input vector $\boldsymbol{X}$ is non-deterministic, i.e., has the probability density $S_0(\boldsymbol{X})$ regardless of uncertainties. Uncertainties usually manifest in a way that merely changes this distribution. The concept of importance sampling can be used to gain insight into dose realizations for different input distributions $S_i(\boldsymbol{X}|\boldsymbol{Z}=\boldsymbol{z_i})$ corresponding to error realizations $\boldsymbol{z}_i, \, i=1,...,N$ from $p_{Z}$, from one initial simulation.
	
	We therefore now recall the basic principles of importance sampling.
	
	\subsection{Standard importance sampling}
	\label{sec:importanceSampling}
	
	Importance sampling is a method originally introduced by \cite{kahnRandomSamplingMonte1950} and later made popular in a more general context by \cite{hastingsMonteCarloSampling1970}. 
	
	Let's assume we are interested in $g(\boldsymbol{X})$. Let $\boldsymbol{X}$ have the probability density function $p(\boldsymbol{X})$ with support $\mathcal{D}_p$, i.e., $\boldsymbol{X} \sim p(\boldsymbol{X})$. Then
	
	\begin{equation}
	\int_{\mathcal{D}_p} g(\boldsymbol{X})p(\boldsymbol{X}) d\boldsymbol{X} = \int_{\mathcal{D}_p} g(\boldsymbol{X}) \frac{p(\boldsymbol{X})}{q(\boldsymbol{X})} q(\boldsymbol{X}) d\boldsymbol{X} \;.
	\end{equation}
	
	So the integral can be approximated as
	
	\begin{equation}
	\int_{\mathcal{Q}} g(\boldsymbol{X}) \frac{p(\boldsymbol{X})}{q(\boldsymbol{X})} q(\boldsymbol{X}) d\boldsymbol{X} \approx \sum_{i=1}^{N} g(\boldsymbol{x_i})  \frac{p(\boldsymbol{x_i})}{q(\boldsymbol{x_i})} \;.
	\end{equation}
	
	Where $\boldsymbol{x_i}$, $i=1,...,N$ are realizations of $\boldsymbol{X}\sim q(\boldsymbol{X})$ and $q$ is a distribution with support $\mathcal{Q} \supseteq \mathcal{D}_p$. 
	
	Thus, the computation of $g(\boldsymbol{X})$ for realizations of $p(\boldsymbol{X})$, can be replaced by computations of $g(\boldsymbol{X})$ with realizations  $\boldsymbol{x_i}$, $i=1,...,N$ from $q(\boldsymbol{X})$ using additional weights $\frac{p(\boldsymbol{x_i})}{q(\boldsymbol{x_i})}$ in the aggregation of the results. This can be convenient if it is easier to sample from $q(\boldsymbol{X})$ than $p(\boldsymbol{X})$ or because choosing $q(\boldsymbol{X})$ cleverly can reduce the variance of the estimate. It can also be helpful, when evaluating $g$ is much more expensive than evaluating $\frac{p(\boldsymbol{x_i})}{q(\boldsymbol{x_i})}$, for example because $g$ is an expensive simulation. In this case, if we are interested in $g(\boldsymbol{X})$ for different distributions of $\boldsymbol{X}$, we can reuse the realizations $g(\boldsymbol{x_1}),...,g(\boldsymbol{x_N})$.
	
	\subsection{Importance (re-)weighting}
	
	For our purpose, this can be applied to mimic dose computations at different points in the parameter space from a previously computed sample of particle trajectories. In the following, we will assume random samples, the method can however be analogously applied to quadratures or quasi-random numbers.
	
	Instead of solving the transport problem $N$ times with slightly different input distributions $S_1,...,S_N$ (compare \cref{eq:doserealizationMC}), it is solved once and the results for other error realizations are determined by (re-)weighting the deposited dose of this one sample of particles as explained in \cref{sec:importanceSampling}. For an error realization $\boldsymbol{z_i}$ from $p_Z$, initial parameter realizations $\boldsymbol{x_1},...,\boldsymbol{x_H}$ from the sampling distribution $q(\boldsymbol{X})$ and target distributions $S_i(\boldsymbol{X}|\boldsymbol{Z}=\boldsymbol{z_i})$, the dose is given by 
	\begin{align}
	\label{eq:IRWsingleReal}
	\nonumber \mathcal{D}_i(\boldsymbol{r}) &= \int_{\mathcal{D}_{S_i}} BB(\boldsymbol{X}(\boldsymbol{z_i})) S_i(\boldsymbol{X}|\boldsymbol{Z}=\boldsymbol{z_i}) d\boldsymbol{X} = \int_{\mathcal{D}_{S_i}} \frac{BB(\boldsymbol{X}(\boldsymbol{z_i}))\;S_i(\boldsymbol{X}|\boldsymbol{Z}=\boldsymbol{z_i})}{q(\boldsymbol{X})}\; \cdot q(\boldsymbol{X}) d\boldsymbol{X}\\ &\approx \frac{1}{H} \sum_{p=1}^{H}  \frac{BB(\boldsymbol{x_p})\;S_i(\boldsymbol{x_p}|\boldsymbol{Z}=\boldsymbol{z_i})}{q(\boldsymbol{x_p})} \;.
	\end{align}
	This yields the following procedure:
	\begin{algorithm}[H]
		\caption{Calculate doses $\mathcal{D}_{S_i}$ }
		\label{alg:IRW_postprocess_vs_onthefly}
		 \begin{multicols}{2}
		\begin{algorithmic} 
			\STATE(a) On-the-fly
			\STATE
			\FOR{particles p=1:H} 
			\STATE Sample $\boldsymbol{x_p} \leftarrow q(\boldsymbol{X})$
			\STATE Compute  $BB(\boldsymbol{x_p})$
			\STATE Sample $\boldsymbol{z_i} \leftarrow p_Z(\boldsymbol{Z})$
			\FOR{target distributions $S_i(\boldsymbol{X}|\boldsymbol{Z}=\boldsymbol{z_i})  \;, \newline i=1,...,N$}
			\STATE $\mathcal{D}_{S_i} \mathrel{+}= \frac{1}{H} \frac{BB(\boldsymbol{x_p})\;S_i(\boldsymbol{x_p}|\boldsymbol{Z}=\boldsymbol{z_i})}{q(\boldsymbol{x_p})}$
			\ENDFOR
			 \ENDFOR
		\end{algorithmic} 
	   \columnbreak
	   		\begin{algorithmic} 
			\STATE(b) In post-processing
			\STATE
			\FOR{particles p=1:H} 
			\STATE Sample $\boldsymbol{x_p} \leftarrow q(\boldsymbol{X})$
			\STATE Compute  $BB(\boldsymbol{x_p})$
			\ENDFOR
			\STATE Sample $\boldsymbol{z_i} \leftarrow p_Z(\boldsymbol{Z})$
			\FOR{target distributions $S_i(\boldsymbol{X}|\boldsymbol{Z}=\boldsymbol{z_i}) \;,\newline i=1,...,N$}
			\STATE $\mathcal{D}_{S_i} \mathrel = \frac{1}{H} \sum_{p=1}^{H} \frac{BB(\boldsymbol{x_p})\;S_i(\boldsymbol{x_p}|\boldsymbol{Z}=\boldsymbol{z_i})}{q(\boldsymbol{x_p})}$
			\ENDFOR
		\end{algorithmic} 
	 \end{multicols}
	\end{algorithm}
	
	The importance and target distributions have to be chosen according to the quantities of interest. Here we show the derivation for the nominal dose without uncertainty, the expected dose and the dose variance with respect to uncertainties. 
	
	\subsection{Nominal dose}
	\label{sec:CompNom}
	
	The nominal dose is the solution to the problem introduced in \cref{sec:radiationTransport} without uncertainty, i.e., for the input distribution $S_0(\boldsymbol{X})$. Let $q(\boldsymbol{X})$ be the simulated particle distribution, then the nominal dose can be computed by replacing $p(\boldsymbol{X})$ with $S_0(\boldsymbol{X})$ in \cref{sec:importanceSampling}.
	
	\begin{align}
	D(\boldsymbol{r}) &= E_{S_0}[BB(\boldsymbol{X})]=\int_{\mathcal{D}_{S_0}} BB(\boldsymbol{X}) S_0(\boldsymbol{X})d\boldsymbol{X} = \int_{\mathcal{D}_{S_0}} \frac{BB(\boldsymbol{X})S_0(\boldsymbol{X})}{q(\boldsymbol{X})}\; \cdot q(\boldsymbol{X}) d\boldsymbol{X} \\
	& \approx \frac{1}{H} \sum_{p=1}^{H}  \frac{BB(\boldsymbol{x_p})\;S_0(\boldsymbol{x_p})}{q(\boldsymbol{x_p})} \nonumber 
	\end{align}
	
	This reduces to the regular dose computation with Monte Carlo described in \cref{sec:radiationTransport}, when $q$ is chosen to be $S_0$.
	
	\subsection{Expected value}
	\label{sec:CompExp}
	
	Using \cref{eq:IRWsingleReal} to compute the dose for individual error realizations, we can mimic the Monte Carlo method for uncertainties and compute the expected dose as 
	
	\begin{align}
	\label{eq:doserealizationsIW}
	\mathcal{D}(\boldsymbol{r})  \approx \frac{1}{N}\sum_{i=1}^N D_i(\boldsymbol{r}) 
	\approx \frac{1}{N}\sum_{i=1}^N \frac{1}{H} \sum_{p=1}^{H}  \frac{BB(\boldsymbol{x_p})\;S_i(\boldsymbol{x_p}|\boldsymbol{Z}=\boldsymbol{z_i})}{q(\boldsymbol{x_p})} \;,
	\end{align}
	
	with $\boldsymbol{z_i}, \, i=1,...,N$ sampled from $S_i$ and $BB(\boldsymbol{x_p})$ again the pre-computed particle trajectories with input parameters $\boldsymbol{x_p}, \, p=1,...,H$ sampled from $q$.
	
	For a known joint distribution of the uncertain input, it is possible to directly compute the expected value, without sampling from $p_Z$, by adapting \cref{eq:expectedDoseJoint}
	
	\begin{equation}
	\label{eq:IRWexpectedDose}
	\mathcal{D}(\boldsymbol{r}) = \mathbb{E}_{\mathcal{S}}[BB(\boldsymbol{X}(\boldsymbol{Z}))] \approx \frac{1}{H} \sum_{p=1}^{H}  \frac{BB(\boldsymbol{x_p})\;\mathcal{S}(\boldsymbol{x_p})}{q(\boldsymbol{x_p})} \;,\end{equation}
	where $\boldsymbol{x_p}, \; p=1,...,H$ sampled from $\mathcal{S}$.
	\subsection{Variance}
	\label{sec:CompVar}
	
	The variance can also be computed for error realizations $\boldsymbol{z_1},...,\boldsymbol{z_N}$ from $p_Z$, using dose realizations from \cref{eq:doserealizationsIW} with the sample variance formula
	
	\begin{equation} 
	\label{eq:IRWvariance}
	\sigma^2 =Var_{p_{\boldsymbol{Z}}}(D(\boldsymbol{r},\boldsymbol{Z}))\approx \frac{1}{N-1}\sum_{i=1}^N \left(D_i(\boldsymbol{r})-\mathcal{D}(\boldsymbol{r})\right)^2  \; .
	\end{equation}

	\subsection{Choosing the importance distribution}
	
	The optimal choice of $q(\boldsymbol{X})$ depends largely on the target quantities. It follows from \cref{sec:CompNom,sec:CompExp,sec:CompVar} that the nominal dose, expected dose and the error scenarios required for the variance computation correspond to $N+2$ different distributions of the input parameters. Thus, when all of these quantities are computed from the same set of realizations, $q(\boldsymbol{X})$ needs to be suitable for not just one but $N+2$ different target distributions. Therefore it will most likely not be possible to choose a function which is optimal for each individual computation and the choice depends on how the accuracy of the different target quantities is prioritized.
	
	For a given target distribution $p(\boldsymbol{X})$ there are two criteria, by which $q(\boldsymbol{X})$ should be chosen (see e.g. \cite{owenMonteCarloTheory2013}):
	\begin{itemize}
		\item Let $Q=\{\boldsymbol{X}|q(\boldsymbol{X}) > 0\}, \text{ if } \; p(\boldsymbol{X}) \cdot BB(\boldsymbol{X})\neq 0 \text{ then } \boldsymbol{X} \in Q$
		\item To minimize the standard error of the approximation $\sigma_q^2=\frac{1}{H}\sum_{p=1}^{H}  \left[ \frac{p(\boldsymbol{x_p})\cdot BB(\boldsymbol{x_p})}{q(\boldsymbol{x_p})}-\mu \right]^2$ with $\mu=E_{S_0}[BB(\boldsymbol{X})]$, $q(\boldsymbol{X})$ should be chosen to be proportional to $p(\boldsymbol{X})\cdot  BB(\boldsymbol{X})$
	\end{itemize}
	
	The first point is jointly attainable for all target values, for example if the support of $q(\boldsymbol{X})$ includes the supports of $S_1(\boldsymbol{X}),...,S_N(\boldsymbol{X})$ as well as those of $S_0$ and $\mathcal{S}$.
	
	The second point creates a trade-off between the different targets. Choosing $q(\boldsymbol{X})=\mathcal{S}$ or $q(\boldsymbol{X})=S_0$, leads to accurate estimates of the expected dose and nominal dose respectively. It can be expected, that $q(\boldsymbol{X})=\mathcal{S}$ is a better choice with respect to the dose variance, since it has more density in the outer regions, which is necessary for an acceptable estimate of the shifted densities $S_i$ for larger errors $\boldsymbol{z_i}$. 
	
	When prioritizing the computation of the variance estimate, a more fat-tailed choice of distribution for $q(\boldsymbol{X})$ is beneficial.  This ensures a larger number of realizations and therefore higher accuracy in the outer regions of the densities, where the variance is typically high. In general it is not advisable to choose a $q$ which is more light-tailed than the target distribution, as the variance can explode if $q$ is too small in regions where the target distribution is not zero \cite{owenMonteCarloTheory2013}.  One option is using a mixture distribution of the individual shifted distributions we want to reconstruct \cite{hesterbergWeightedAverageImportance1995}. Another common rule of thumb in case of a normal target distribution is to use a Student's t distribution, as they are similar but the latter has heavier tails \cite{owenMonteCarloTheory2013}. The accuracy of estimates for some exemplary choices of $q$ will be investigated numerically in \cref{sec:numError}.
	
	\section{Mathematical properties}
	\label{sec:mathematicalProperties}
	\subsection{Accuracy \& Bias}
	An advantage of the proposed method is that it inherits some of the mathematical properties of importance sampling. Namely, for the nominal dose, expected dose and dose for the single error realizations, the estimates are unbiased \cite{owenMonteCarloTheory2013}. 
	\begin{proposition}
	 For $q(\boldsymbol{X}): \mathcal{D}_X \rightarrow \mathbb{R}_+$ with $Q=\{\boldsymbol{X}|q(\boldsymbol{X}) > 0\}, \text{ if } \; p(\boldsymbol{X}) \cdot BB(\boldsymbol{X})\neq 0 \text{ let } \boldsymbol{X} \in Q$ . Then the importance reweighting estimator $\hat{D}(\boldsymbol{r})$ is unbiased, i.e. $E_q[\hat{D}(\boldsymbol{r})] = D(\boldsymbol{r})$.
	 \label{prop:1}
	\end{proposition}
	Proof:
	According to \cref{sec:CompNom}, the importance sampling estimate is defined as 
	\begin{equation}\hat{D}(\boldsymbol{r})=\frac{1}{H} \sum_{p=1}^{H}  \frac{BB(\boldsymbol{x_p})\;p(\boldsymbol{x_p})}{q(\boldsymbol{x_p})} \;,\end{equation}
	where $\boldsymbol{x_p},\;p=1,...,H$ are realizations of $\boldsymbol{X}\sim q(\boldsymbol{X})$. Then
	\begin{align}
	E_q[\hat{D}(\boldsymbol{r})] &= E_q\left[\frac{1}{H} \sum_{p=1}^{H}  \frac{BB(\boldsymbol{x_p})\;p(\boldsymbol{x_p})}{q(\boldsymbol{x_p})} \right] = \frac{1}{H} \sum_{p=1}^{H} E_q\left[\frac{BB(\boldsymbol{x_p})\;p(\boldsymbol{x_p})}{q(\boldsymbol{x_p})}\right] \\
	&= \int_{\mathcal{Q}} \frac{BB(\boldsymbol{X})\;p(\boldsymbol{X})}{q(\boldsymbol{X})} \cdot q(\boldsymbol{X}) d\boldsymbol{X} = \int_{\mathcal{Q}} BB(\boldsymbol{X}) \cdot p(\boldsymbol{X}) d\boldsymbol{X} \;, \nonumber \\
	&= \int_{\mathcal{D}_{p}} BB(\boldsymbol{X}) \cdot p(\boldsymbol{X}) d\boldsymbol{X} + \int_{\mathcal{Q}\cap\mathcal{D}_{p}^c } \underbrace{BB(\boldsymbol{X})}_{=0} \cdot p(\boldsymbol{X}) d\boldsymbol{X} +\int_{\mathcal{Q}^c \cap \mathcal{D}_{p}} BB(\boldsymbol{X}) \cdot \underbrace{p(\boldsymbol{X})}_{=0} d\boldsymbol{X} \;,\nonumber\\
	& = E_{S_0}[BB(\boldsymbol{X})]=D(\boldsymbol{r}) \;\nonumber
	\end{align}\qed.
	
	Also, the standard error $\epsilon_q$ of these estimators can be computed similarly to that of Monte Carlo estimates \cite{owenMonteCarloTheory2013}. It depends on the number of realizations as well as the similarity of the importance distribution and the distribution of interest:
	
	\begin{equation}
	\label{TheoreticalError}
	\epsilon_q = \frac{\sigma_q^2}{H} \,\text{, where  }  \sigma_q^2 \approx
	\hat\sigma_q^2=\frac{1}{H-1}\sum_{p=1}^{H}  \left[\frac{BB(\boldsymbol{x_p})S_0(\boldsymbol{x_p})}{q(\boldsymbol{x_p})}-\hat{D}(\boldsymbol{r})\right]^2 \;. \end{equation}

	To lower the standard error, the importance distribution should resemble the target distribution. In the variance computation, the estimates for more extreme error realizations will thus be less accurate. For the type and extent of errors in radiation therapy, the distributions are however mostly similar enough, as we will demonstrate further in \cref{sec:results}.
	
	In contrast to the dose estimates, the variance computed as in \cref{sec:CompVar} is not generally unbiased. This is due to its computation from estimates of the dose for shifted beams, which rely on the same single sample of particles. What we consider as the realizations in the sample variance computation, while individually unbiased, are thus no longer independent. 
	\begin{proposition}
	For $\boldsymbol{x_p}$ from $S_i(\boldsymbol{x_p}|\boldsymbol{Z} = \boldsymbol{z_i})$, let $w_i(\boldsymbol{x_p})=\frac{S_i(\boldsymbol{x_p}|\boldsymbol{Z} = \boldsymbol{z_i})}{q(\boldsymbol{x_p})}$ be independent of $BB(\boldsymbol{x_p})$. Then the importance reweighting estimator for the dose variance is not generally unbiased and the bias is given by \[b_{S_N^2} = \left( \frac{1}{H} \sum_{p=1}^H BB(\boldsymbol{x_p}) \cdot E[w_i(\boldsymbol{x_p})] \right)^2 -\frac{1}{H^2}\sum_{p=1}^H \sum_{h=1}^H BB(\boldsymbol{x_p})BB(\boldsymbol{x_h}) \cdot E[w_i(\boldsymbol{x_p})w_j(\boldsymbol{x_h})]\] 
	\label{prop:2}
	\end{proposition}
	
	Proof: In \cite{benhamouFewPropertiesSample2018}, general formulas for the moments of the sample variance $S_n^2$ are derived. The first moment $E[S_n^2]$ for general non independent or identically distributed (iid) samples $X_1,...,X_N$ is given by:
	\begin{equation}
	E[S_n^2] = \frac{\sum_{i=1}^n \mathbb{E}[X_i^2]}{n}  -\frac{\sum_{i \neq j} \mathbb{E}[X_iX_j]}{n(n-1)} \; X_1,...,X_n \text{not iid} \nonumber\;.
	\end{equation}
	
	Applied to our context,  the realizations are $D_i= \frac{1}{H}\sum_{p=1}^{H} BB(\boldsymbol{x_p})w_i(\boldsymbol{x_p})$ for $i=1,...,N$, with $w_i(\boldsymbol{x_p}) = \frac{S_i(\boldsymbol{x_p}|\boldsymbol{Z} = \boldsymbol{z_i})}{q(\boldsymbol{x_p})}$. Since all realizations use the same sample of computed particle trajectories $BB(\boldsymbol{x_p})$ and initial particle properties $\boldsymbol{x_p}, \; p=1,...,H$, they cannot be assumed to be independent, in particular $D_i = D_j + \frac{1}{H} \sum_{p=1}^H BB(\boldsymbol{x_p}) \left( w_i(\boldsymbol{x_p}- w_j(\boldsymbol{x_p})\right)$. Thus,
	
	\begin{align}
	E[S_n^2] &= \frac{\sum_{i=1}^N E[D_i^2]}{N}  -\frac{\sum_{i \neq j} E[D_iD_j]}{N(N-1)} \;.
	\end{align}
	Since we assume that the $D_i$ are identically distributed, this reduces to
	
	\begin{align}
	 	E[S_n^2] &= E[D_i^2] - E[D_iD_j]  \nonumber\\
	&= \underbrace{E[D_i^2] - E[D_i]E[D_j]}_{=Var(D)} - \underbrace{\left( E[D_i]E[D_j] - E[D_iD_j] \right)}_{=b_{S_N^2}\neq 0\;, \text{ for} \;i \neq j}   \;.\nonumber\\ 
	\end{align}
	
	Thus the estimator is not necessarily unbiased and the bias is
	\begin{align}
	b_{S_N^2} &= \underbrace{E[D_i]E[D_j]}_{=E[D_i]E[D_i]} - E[D_iD_j]   \nonumber\\
	&= E \left[ \frac{1}{H}\sum_{p=1}^{H} BB(\boldsymbol{x_p})w_i(\boldsymbol{x_p}) \right]^2 - E \left[ \left( \frac{1}{H}\sum_{p=1}^{H} BB(\boldsymbol{x_p})w_i(\boldsymbol{x_p}) \right)\left( \frac{1}{H}\sum_{p=1}^{H} BB(\boldsymbol{x_p})w_j(\boldsymbol{x_p}) \right) \right] \nonumber\\
	&=   \left(\frac{1}{H}\sum_{p=1}^{H} BB(\boldsymbol{x_p})E \left[w_i(\boldsymbol{x_p}) \right]\right)^2 - E \left[  \frac{1}{H^2}\sum_{p=1}^{H}\sum_{h=1}^{H} BB(\boldsymbol{x_p})BB(\boldsymbol{x_h})w_i(\boldsymbol{x_p})w_j(\boldsymbol{x_h}) \right] \nonumber\\
	&=  \left( \frac{1}{H}\sum_{p=1}^{H} BB(\boldsymbol{x_p})E \left[w_i(\boldsymbol{x_p}) \right] \right) ^2 -  \frac{1}{H^2}\sum_{p=1}^{H}\sum_{h=1}^{H} BB(\boldsymbol{x_p})BB(\boldsymbol{x_h})E \left[ w_i(\boldsymbol{x_p})w_j(\boldsymbol{x_h}) \right] \end{align} \qed.
	
    Since this bias term is costly to compute, it is generally not feasible to score this in addition to the quantities of interest. Computing the bias once for a given problem class could however be of interest to judge its order of magnitude. Its potential impact on our results will be assessed in greater detail in \cref{sec:results}.
	
	\subsection{Bounds on the variance}
	\label{sec:varianceBound}
	While the discussed estimators can be expected to significantly lower the expense of scenario computations, the variance still requires $N$ scoring operations, either accumulated on-the-fly after each simulated particle history, or in post processing as (re-)weighting of stored histories. For a large number of simulated particles, run-times can therefore exceed the required time-frame, especially in the context of robust optimization, where the variance has to be computed in each of numerous iterations \cite{pflugfelderWorstCaseOptimization2008, liuRobustOptimizationIntensity2012, chuRobustOptimizationIntensity2005}.
	\begin{proposition}
	An upper bound to the variance estimator $\hat{\sigma}^2$ is given by \[ \hat{\sigma}^2 \leq \left(\; \frac{1}{H} \sum_{p=1}^{H} BB(\boldsymbol{x_p})^2 \;\right) \left(\; \frac{1}{H} \sum_{p=1}^{H} S_N^2(w(\boldsymbol{x_p})) \; \right)\]
	\label{prop:3}
	\end{proposition}
	
	Proof: Using the definition of the importance reweighing estimator for the dose variance, the dose for a single error realization and the expected dose (see \cref{eq:IRWvariance,eq:IRWsingleReal,eq:IRWexpectedDose}), we get
	\begin{align}
	 \hat{\sigma}^2	 &= \frac{1}{N-1} \sum_{i=1}^{N} \; \left[ \; \frac{1}{H} \sum_{p=1}^{H}  \frac{BB(\boldsymbol{x_p})\;S_i(\boldsymbol{x_p}|\boldsymbol{Z}=\boldsymbol{z_i})}{q(\boldsymbol{x_p})} - \frac{1}{N}\sum_{j=1}^N \frac{1}{H} \sum_{p=1}^{H}  \frac{BB(\boldsymbol{x_p})\;S_j(\boldsymbol{x_p}|\boldsymbol{Z}=\boldsymbol{z_j})}{q(\boldsymbol{x_p})}\; \right]^2 \nonumber \\
	 &= \frac{1}{N-1} \sum_{i=1}^{N} \; \left[ \; \frac{1}{H} \sum_{p=1}^{H}  BB(\boldsymbol{x_p})\;w_i(\boldsymbol{x_p}) - \frac{1}{N}\sum_{j=1}^N \frac{1}{H} \sum_{p=1}^{H}  BB(\boldsymbol{x_p})\;w_j(\boldsymbol{x_p})\; \right]^2 \nonumber \\
	 &= \frac{1}{N-1} \sum_{i=1}^{N} \; \left[ \; \frac{1}{H} \sum_{p=1}^{H}  BB(\boldsymbol{x_p})\;w_i(\boldsymbol{x_p}) - \frac{1}{H} \sum_{p=1}^{H}  BB(\boldsymbol{x_p}) \frac{1}{N}\sum_{j=1}^N \;w_j(\boldsymbol{x_p})\; \right]^2 \nonumber \\
	 &= \frac{1}{N-1} \sum_{i=1}^{N} \; \left[ \; \frac{1}{H} \sum_{p=1}^{H}  BB(\boldsymbol{x_p}) \left( w_i(\boldsymbol{x_p}) -  \frac{1}{N}\sum_{j=1}^N \;w_j(\boldsymbol{x_p}) \right)\; \right]^2 \nonumber \\
	&\leq\left(\; \frac{1}{H} \sum_{p=1}^{H} BB(\boldsymbol{x_p})^2\right)  \;\Biggl(\frac{1}{H} \sum_{p=1}^{H} \; \underbrace{\frac{1}{N-1} \sum_{i=1}^{N} \left[ w_i(\boldsymbol{x_p})- \sum_{j=1}^{N} w_j(\boldsymbol{x_p}) \right]^2}_{= \, S_N^2(w(\boldsymbol{x_p}))\, \approx \, Var(w(\boldsymbol{x_p}))} \Biggr) \nonumber\\\
	&= \left(\; \frac{1}{H} \sum_{p=1}^{H} BB(\boldsymbol{x_p})^2 \;\right) \left(\; \frac{1}{H} \sum_{p=1}^{H} S_N^2(w(\boldsymbol{x_p})) \; \right) \; \nonumber
	\end{align}\qed.
	
	The advantage of this bound is, that the squared simulation response $BB(\boldsymbol{x_p})^2 $ is aggregated independently from the error realizations. This includes the implicit assumption, that the variance in the likelihood ratio between the simulated and target distribution, is enough to describe the variance of the output. While this might hold for relatively homogeneous background media, it is unclear whether it is sufficient in heterogeneous media as present in many patient anatomies.

	\section{Modeling set-up uncertainties}
	\label{sec:modelingUncertainties}
	
	We now consider an application of the proposed approach to the case of uncertainties in the patient set-up or position.
	These can be modeled by random shifts in the particle positions $\boldsymbol{r}$. The input $\boldsymbol{X}$ is thus affected by the error additively, i.e., the input $\boldsymbol{X(Z)}$ under uncertainty is
	\begin{equation}
	\boldsymbol{X(Z)} = \boldsymbol{X} + \boldsymbol{Z} \;, \text{where } \boldsymbol{X}=(\boldsymbol{r},\boldsymbol{\Omega},E) \text{ and } \boldsymbol{Z}=(\boldsymbol{Z_r},\boldsymbol{0},0).
	\end{equation}
	
	We follow the common assumption, that errors as well as input parameters are normally distributed	\cite{wieserImpactGaussianUncertainty2020,unkelbachAccountingRangeUncertainties2007,perkoFastAccurateSensitivity2016, bangertAnalyticalProbabilisticModeling2013, fredrikssonMinimaxOptimizationHandling2011, wahlEfficiencyAnalyticalSamplingbased2017}
	\begin{equation}
	p_{Z}=\mathcal{N}(\boldsymbol{\mu}_{Z},\boldsymbol{\Sigma}) \;, \; S_0=\mathcal{N}(\boldsymbol{\mu}_X,\boldsymbol{\Lambda}) \;.
	\end{equation}
	
	In this case we can define the density $\mathcal{S}$ of the input including the uncertain factors through convolution of $p_{Z}$ and $S_0$:
	
	\begin{equation}
	\mathcal{S} := \mathcal{N}(\boldsymbol{\mu}_Z+\boldsymbol{\mu}_{X},\boldsymbol{\Sigma}+\boldsymbol{\Lambda}) \;.
	\end{equation}
	
	Thus, the expected value can be directly computed using \cref{sec:CompExp} and the nominal dose as well as the dose variance are computed as described in \cref{sec:CompNom,sec:CompVar}.

	\subsection{Beam correlation models}
	
	The most frequently used uncertainty model for set-up errors assumes one global error $\boldsymbol{Z_r}=(Z_{r_x}, Z_{r_y}, Z_{r_z}) \sim \mathcal{N}(\boldsymbol{\mu}_{Z_r},\boldsymbol{\Sigma}_r).$ Therefore, it corresponds to movements or mispositioning of the patient which affect the complete body or at least area of interest around the tumor. 
	
	However, in radiation therapy, often numerous beamlets with different energy levels, positions and incoming angles are used in order to achieve better coverage of the tumor. Here, the term \textit{beamlet} or \textit{pencil beam} refers to a single instance of energy, angular and lateral position. An ensemble of beamlets coming from the same geometrical set-up of nozzle and patient is then referred to as \textit{beam}, a group of beamlets with the same lateral positioning within a beam as \textit{ray} and a group of beamlets with the same energy within a beam as \textit{energy level}. In this case, it may be more reasonable to assume beamlet-specific correlation patterns, e.g., depending on the differences in time and position of application of the individual beamlets. Additionally, treatments are usually administered in several sittings or fractions, which can also affect how errors are correlated. A number of simple correlation models, assuming full correlation between the beamlets belonging to the same beam or ray, have been investigated by different authors \cite{bangertAnalyticalProbabilisticModeling2013, wahlEfficiencyAnalyticalSamplingbased2017, pflugfelderWorstCaseOptimization2008} and also discussed in our previous work \cite{stammerEfficientUncertaintyQuantification2021}. However, depending on the dose calculation and uncertainty quantification method, the incorporation of more complex correlation models can be tedious to both set up and compute.
		
	Fortunately, the introduced framework principally allows arbitrary correlation models for uncertainties in the set-up of different beamlets. These can be implemented using the covariance matrix of a multivariate error distribution. We define the errors in each beamlet by $\boldsymbol{Z^b_r}=(Z^b_{r_x}, Z^b_{r_y}, Z^b_{r_z}) \sim \mathcal{N}(\boldsymbol{\mu}^b_{Z_r},\boldsymbol{\Sigma}_r^b)$ and their correlations by
	
	\begin{equation}  C = \begin{pmatrix}
	\Sigma_r^1 & \begin{matrix}
	\rho_{xx}^{12} & \rho_{xy}^{12} & \rho_{xz}^{12} \\
	\rho_{yx}^{12} & \rho_{yy}^{12}  & \rho_{yz}^{12} \\
	\rho_{zx}^{12} & \rho_{zy}^{12}  & \rho_{zz}^{12} \\
	\end{matrix}  & \cdots & \begin{matrix}
	\rho_{xx}^{1B} & \rho_{xy}^{1B} & \rho_{xz}^{1B} \\
	\rho_{yx}^{1B} & \rho_{yy}^{1B}  & \rho_{yz}^{1B} \\
	\rho_{zx}^{1B} & \rho_{zy}^{1B}  & \rho_{zz}^{1B} \\
	\end{matrix} \\
	
	\begin{matrix}
	\rho_{xx}^{21} & \rho_{xy}^{21} & \rho_{xz}^{21} \\
	\rho_{yx}^{21} & \rho_{yy}^{21}  & \rho_{yz}^{21} \\
	\rho_{zx}^{21} & \rho_{zy}^{21}  & \rho_{zz}^{21} \\
	\end{matrix} & \Sigma_r^2 & & \\
	
	\vdots & &\ddots & \vdots \\
	
	\begin{matrix}
	\rho_{xx}^{B1} & \rho_{xy}^{B1} & \rho_{xz}^{B1} \\
	\rho_{yx}^{B1} & \rho_{yy}^{B1}  & \rho_{yz}^{B1} \\
	\rho_{zx}^{B1} & \rho_{zy}^{B1}  & \rho_{zz}^{B1} \\
	\end{matrix} & \cdots & &\Sigma_r^B \end{pmatrix} \;.
	\end{equation}
	Here $\rho_{xy}^{ab}$ is the correlation between the set-up errors of beamlet $a$ in dimension $x$ and beamlet $b$ in dimension $y$. It is then possible to obtain error realizations for different choices of the correlations, using a known probability distribution or, alternatively, a Copula with mean $\boldsymbol{\mu}_r=(\boldsymbol{\mu}^1_{Z_r},...,\boldsymbol{\mu}^B_{Z_r})^T$ and covariance matrix $\boldsymbol{C}$.
	
	Besides the discussed simple correlation models, this framework allows us to model more complex correlation patterns. This can, for example, be used to incorporate knowledge about the radiation process and even model certain time dependencies. In the following, we demonstrate the derivation of the covariance matrix $\boldsymbol{C}$ for two examples of autocorrelated set-up errors.
	
	\subsubsection{Autocorrelated set-up errors}
	
	To show how the covariance matrix $\boldsymbol{C}$ can be obtained for a given correlation model, we first consider the case of autocorrelated set-up errors, which follow an AR(1) process. While irradiation itself usually only takes a few milliseconds, beamlets are applied sequentially according to their energy level and up to 1-2 seconds can pass between them. Thus, assuming the patient is exhibiting random movement with some probability at all points in time, the patient position corresponding to beamlets in adjacent energy groups should be correlated to a higher degree than those of beamlets with a larger difference in energies. Since first order autoregressive models are sufficient to capture interfractional set-up errors \cite{huTimeSeriesAnalysis2012}, we choose such a model for the random patient movements during treatment. 
	
	We thus describe movements by an autoregressive process of order one, meaning that the overall deviation at time $t$ depends on the deviation at time $t-1$ as well as a random variable $\boldsymbol{U}_t  \simiid  \mathcal{N}(\boldsymbol{\mu}_{Z_r},\boldsymbol{\Sigma}_r)$, which represents the movement error introduced between $t-1$ and $t$:
	
	\begin{equation}\boldsymbol{Z}_{r;t} = \alpha \boldsymbol{Z}_{r;t-1} + \boldsymbol{U}_t \;.\end{equation}
	
	Here the factor $|\alpha| <1 $ controls how strongly the position at the previous time step affects the current position. It can for example be chosen depending on how much time passes between the time steps. In our case each beam $b$ corresponds to a point in time $t(b)$ according to its energy and we assume that there is no additional correlation between the (x,y,z)-coordinates, i.e., 
	\begin{equation} \boldsymbol{\Sigma}= \begin{pmatrix}
	\boldsymbol{\sigma}_x^2 & 0 & 0 \\
	0 & \boldsymbol{\sigma}_y^2 & 0 \\
	0 & 0 & \boldsymbol{\sigma}_z^2 \\
	\end{pmatrix} \;.\end{equation}
	
	From this description of the process it is now possible to obtain the covariance matrix $\boldsymbol{C}$ using the Autocovariance Function (ACVF):
	
	\begin{equation} \rho_{ab}^{ij} = Cov(Z_{r_a;t(i)},Z_{r_b;t(j)})= \begin{cases} \frac{\alpha^{|t(i)-t(j)|} \sigma_x^2}{1-\alpha^2} \;\;, a=b=x\\
	\\
	\frac{\alpha^{|t(i)-t(j)|} \sigma_y^2}{1-\alpha^2}\;\; ,a=b=y\\
	\\
	\frac{\alpha^{|t(i)-t(j)|} \sigma_z^2}{1-\alpha^2}\;\; ,a=b=z \\ 
	\\
	0 \;\;\;\;\;\; \;\;\;\;\;\; ,a \neq b
	\end{cases} \;.\end{equation}
	
	Note that the variance is constant in time and the covariances only depend on the time difference. To obtain a joint probability function we use a Gaussian copula, which for Gaussian marginals reduces to the multivariate normal distribution $\mathcal{N}(\boldsymbol{\mu}_{Z_r},\boldsymbol{C})$ with  $\boldsymbol{\mu}_{Z_r} =( \boldsymbol{\mu}_{Z_r}^1, \boldsymbol{\mu}_{Z_r}^2, ..., \boldsymbol{\mu}_{Z_r}^B)$ and 
	\footnotesize
	\begin{equation}
	\boldsymbol{C}=\begin{pmatrix}
	\frac{\sigma_x^2}{1-\alpha^2}& 0 & 0 & && & \frac{\alpha^{|t(B)-t(1)|}\sigma_x^2}{1-\alpha^2} & 0 & 0 \\
	0 &\frac{\sigma_y^2}{1-\alpha^2}  & 0 & & \cdots& & 0 & \frac{\alpha^{|t(B)-t(1)|}\sigma_y^2}{1-\alpha^2} & 0 \\
	0 &0 & \frac{ \sigma_z^2}{1-\alpha^2} &&& & 0 & 0 & \frac{\alpha^{|t(B)-t(1)|} \sigma_z^2}{1-\alpha^2} \\
	&&&&&&&& \\
	& \vdots & & &\ddots& & & \vdots & \\
	&&&&&&&& \\
	\frac{\alpha^{|t(1)-t(B)|}\sigma_x^2}{1-\alpha^2}& 0 & 0 & && & \frac{\sigma_x^2}{1-\alpha^2}& 0 & 0 \\
	0 &\frac{\alpha^{|t(1)-t(B)|}\sigma_y^2}{1-\alpha^2}  & 0 &&\cdots& & 0 &\frac{\sigma_y^2}{1-\alpha^2}  & 0 \\
	0 &0 & \frac{\alpha^{|t(1)-t(B)|} \sigma_z^2}{1-\alpha^2} &&& & 0 &0 & \frac{ \sigma_z^2}{1-\alpha^2}\\
	&\end{pmatrix} \;.
	\end{equation}
	\normalsize
	This AR(1)-correlation model gives an intuitive description for random movements of a patient over the time of irradiation. However, there are also errors caused by periodic movements such as respiratory motion or heart beats. Respiratory motion is often modeled using Gaussian processes with an appropriate covariance kernel \cite{durichenMultitaskGaussianProcesses2015}, however, there is also literature exploring the use of a periodic autoregressive moving average (ARMA) model \cite{hommaNewMotionManagement2009}. Both could be incorporated into our uncertainty framework, we however confine ourselves to showing an example for a Gaussian process with a (local) periodic kernel here. 
	
	The Gaussian process (GP) implements the concept of data in a time series following a joint multivariate Gaussian distribution, where the covariance kernel defines the dependencies between different data points. As the type of movement we are interested in has a dominant periodic component, which may however vary over time, we choose the following kernel $k(Z_{r;t(i)},Z_{r;t(j)})$, proposed similarly by \cite{duvenaudAutomaticModelConstruction2014},
	
	\begin{equation}
	\label{LocalPeriodicKernel}
	k(Z_{r;t(i)},Z_{r;t(j)})= \sigma^2 \cdot\exp\left(-\frac{2 \sin^2(\pi(t(i)-t(j))/p)}{l_1^2}\right) \cdot \exp\left(-\frac{(t(i)-t(j))^2}{2l_2^2}\right) \;.
	\end{equation}
	
\begin{figure}
		\centering
		\begin{subfigure}{0.35\textwidth}
		\centering
			\includegraphics[height=4.75cm]{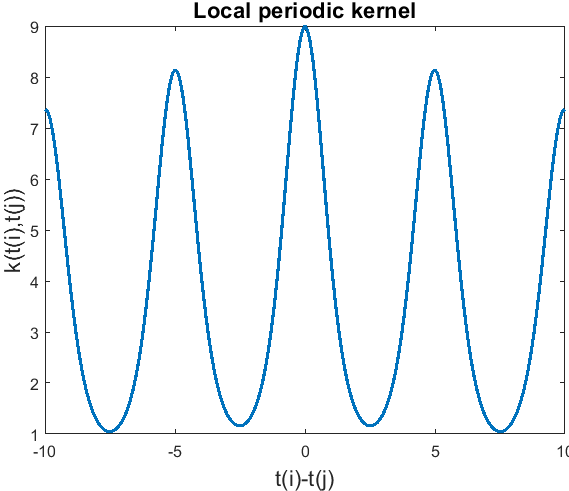}
			\caption*{(a)}
		\end{subfigure}
		\begin{subfigure}{0.6\textwidth}
		\centering
			\includegraphics[height=4.75cm]{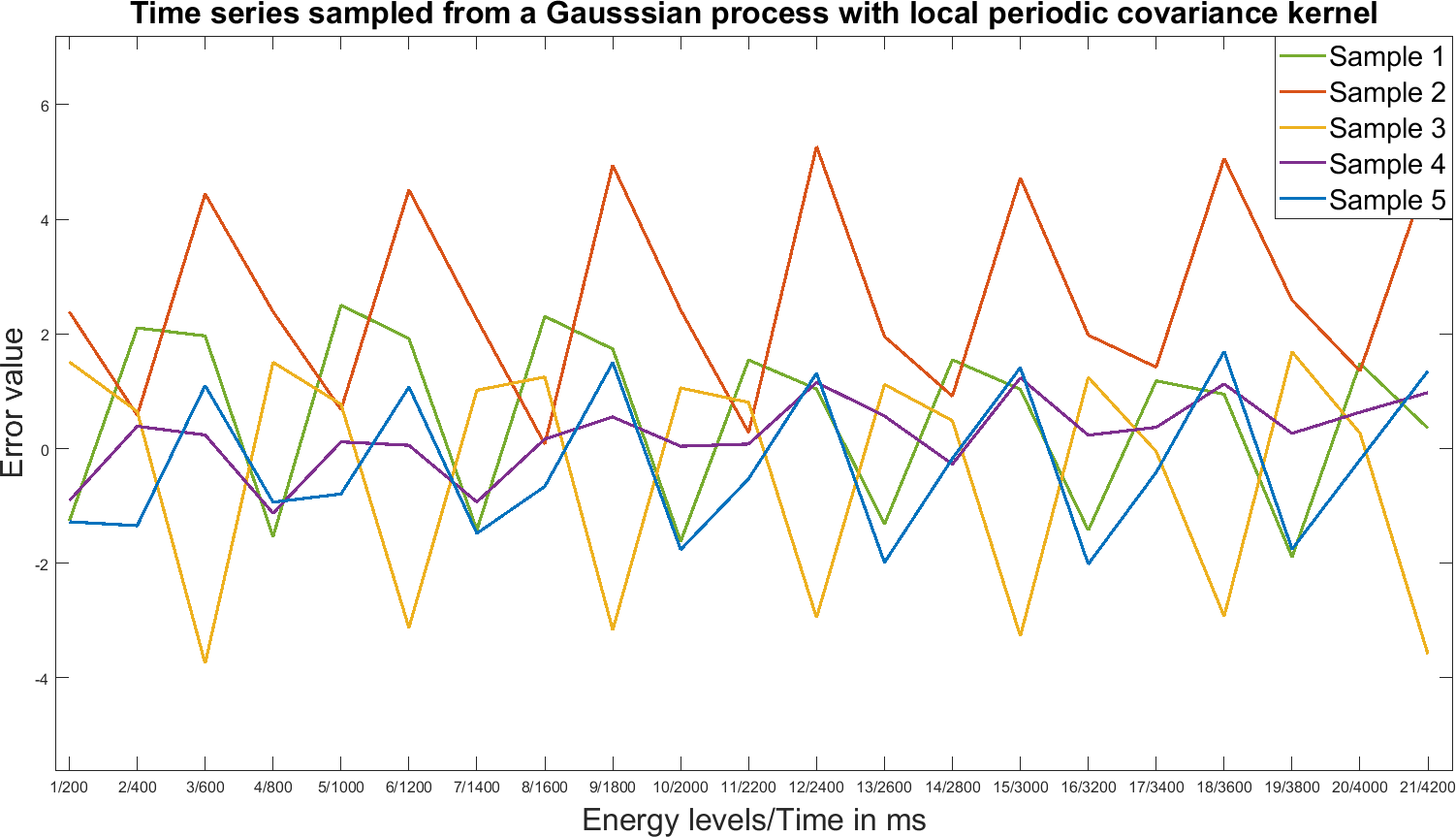}
			\caption*{(b)}
		\end{subfigure}
		\caption{Local periodic covariance kernel with  $\sigma=5$, $p=3$,$l_1=1$ and $l_2=5$ (a) and five realizations of the associated GP (b)}
		\label{fig:LocalPeriodicKernel}
	\end{figure}

	Here $\sigma^2$ is the variance of the stationary process, $p$ is the period and $l_1$ and $l_2$ are scale parameters. The first exponential term describes an exact periodic repetition, whereas the second term is a local kernel which introduces the variations over time. \Cref{fig:LocalPeriodicKernel} illustrates the covariance kernel as well as five exemplary realizations. The process starts at a random point in the cycle and periodically repeats itself with some random variations. From \cref{LocalPeriodicKernel} the derivation of the covariance matrix $\boldsymbol{C}$ is straightforward with the entries
	
	\begin{equation}
	\rho_{ab}^{ij} = Cov(Z_{r_a;t(i)},Z_{r_b;t(j)})= \begin{cases} \sigma_x^2 \cdot\exp(-\frac{2 \sin^2(\pi(t(i)-t(j))/p)}{l_1^2}) \cdot \exp(-\frac{(t(i)-t(j))^2)}{2l_2^2}) \;\;, a=b=x\\
	\\
	\sigma_y^2 \cdot\exp(-\frac{2 \sin^2(\pi(t(i)-t(j))/p)}{l_1^2}) \cdot \exp(-\frac{(t(i)-t(j))^2)}{2l_2^2})\;\; ,a=b=y\\
	\\
	\sigma_z^2 \cdot\exp(-\frac{2 \sin^2(\pi(t(i)-t(j))/p)}{l_1^2}) \cdot \exp(-\frac{(t(i)-t(j))^2)}{2l_2^2})\;\; ,a=b=z \\ 
	\\
	0 \;\;\;\;\;\; \;\;\;\;\;\; ,a \neq b
	\end{cases} \;.
	\end{equation}
	
	The errors for all pencil beams can now be sampled from the multivariate Gaussian $\mathcal{N}(\boldsymbol{\mu}_{\delta},\boldsymbol{C})$ and will depend on the time at which their corresponding energy level is applied. 
	
	The two examples above demonstrate, that it is possible to implement complex movement patterns and even to some degree model time dependencies, by changing the way error realizations are sampled. 
	
	\section{Implementation and Test cases}
	The described method was implemented as post-processing in Matlab. Radiation plans were generated for proton dose calculations using the treatment planning software matRad \cite{wieserDevelopmentOpensourceDose2017} and dose calculations were performed with the Monte Carlo engine TOPAS \cite{perlTOPASInnovativeProton2012}. For quicker convergence we use randomized quasi-Monte Carlo with scrambled Sobol numbers when sampling the error scenarios for the dose variance estimate (see \cref{sec:RQMC}). 
	
	In the following, we present the results for a 3D water phantom as well as a prostate and liver patient obtained from the open CORT (common optimization for radiaion therapy) data set \cite{craftSharedDataIntensity2014}. Overall, we consider three different treatment plans, combined with the three error models discussed in \cref{sec:modelingUncertainties}. For the Gaussian set-up error we assume a standard deviation of $\SI{3}{\milli\meter}$ and mean $\mu_{\delta}=0$, i.e., the errors do not have a systematic component \cite[comp.][]{perkoFastAccurateSensitivity2016}. For two of the treatment plans, intensity modulated proton therapy (IMPT) \cite{kanaiSpotScanningSystem1980, lomaxIntensityModulationMethods1999} was employed. In these cases, each source consists of a large number of narrow pencil beams to enable a better and more flexible coverage of the target volume. More details concerning the treatment plans and the respective error models applied to each of them can be found in \cref{table:patientStatistics}.
	
	\begin{table}[htb!]
		\centering
		\caption{Overview of simulated plans and error models per test case.}
		\begin{tabular}{l c c c c}
			\hline
			Test case & \multicolumn{2}{c}{Water phantom} & Prostate & Liver\\
			\hline
			Size& \multicolumn{2}{c}{60x25x25 voxels} & 183x183x90 voxels & 217x217x168 voxels \\
			Irradiation angles & \multicolumn{2}{c}{(\ang{0},\ang{0})} & (\ang{0},\ang{90})/(\ang{0},\ang{270}) & (\ang{0},\ang{315})\\
			Pencil beams & 1 & 175 & 1$\,$375/1$\,$383  & 1408\\
			Simulated particles & 100$\,$000 & 2$\,$566$\,$453 & 16$\,$992$\,$193/16$\,$748$\,$034 & 13 528 430  \\
			Error model & global & global/AR(1)/GP & global & global/AR(1)/GP \\
			Sample size & 500 & 500 & 100 &100\\
			\hline
		\end{tabular}
		\label{table:patientStatistics}
	\end{table}

	\subsection{Evaluation criteria}
	In order to evaluate the quality of our estimates, we compute a reference for each test case using the regular randomized quasi-Monte Carlo method for the uncertainties. The number of realizations was chosen to be equal to that used for the importance (re-)weighting estimates, which is 500 for the water phantom and 100 for the prostate patient.
	
	We compare the results according to the $\gamma$ criterion \cite{lowTechniqueQuantitativeEvaluation1998}, as well as a difference map. For the difference maps, we compute 
	\begin{equation}
	\text{diff}_i(d^\text{ref}_i,d_i^\text{est}) = d_i^\text{ref} - d_i^\text{est} \; ,
	\end{equation}
	for each voxel $i$ in the reference result $d^\text{ref}$ and (re-)weighting estimate $d^\text{est}$.
	
	The $\gamma$ criterion is frequently used for comparisons of three-dimensional dose matrices in clinical applications. To compare doses $d$ and $d_c$, one first computes
	
	\begin{align}
	\gamma(x) = \min_{x_c} \sqrt{\frac{|x_c - x|^2}{\Delta \tilde{x}^2} + \frac{|d_c(x_c) - d(x)|^2}{\Delta \tilde{d}^2}} \; ,
	\end{align}
	
	where $x$ and $d(x)$ are a location and the dose at that location respectively, $x_c$ and $d_c(x_c)$ are locations and dose in the cube to be compared and $\Delta \tilde{x}$ and $\Delta \tilde{d}$ are the user defined distance and dose tolerances. The overall agreement is then measured according to the $\gamma$-passrate over all points $x$. For each point
	
	\begin{equation}
	\label{GammaCriterion}
	\begin{cases} \gamma(x) \leq 1,\; x \text{ passes} \\
	\gamma(x) > 1, \;x \text{ fails} \; .
	\end{cases}\end{equation}
	
	 It implements the idea of agreement within a predefined neighborhood and tolerance around a single coordinate or data point. In contrast to criteria such as the difference or (root) mean square error, which compare the values at an exact location, this does not take minor fluctuations in the spacial dose distribution into account. Since the reference itself is also subject to statistical errors, the dose distribution is not necessarily incorrect if it deviates slightly from the reference. However, the $\gamma$ criterion is not a metric in the mathematical sense if the tolerances are larger than zero. Thus, two dose cubes could have $100\%$ agreement according to the $\gamma$-passrate without actually being the same. Therefore the $\gamma$-passrates always have to be interpreted together with the applied tolerances. In the following, we choose values frequently used in literature: a distance tolerance of $\SI{3}{\mm}$ and a dose tolerance of $\SI{3}{\percent}$. Furthermore, the $\gamma$-passrate can depend on the grid resolution as well as the implementation \cite{husseinChallengesCalculationGamma2017} . We use the matRad implementation \cite{wieserDevelopmentOpensourceDose2017}, which reduces the dependency on the resolution by interpolating the comparison dose cube and comparing it at a higher resolution.
	
	\section{Results }
	\label{sec:results}

	\subsection{Fully correlated pencil beams}
	
\begin{figure}[H]
		\centering 
		\begin{minipage}{0.05\textwidth}
			\vspace*{\fill}
			$E[\boldsymbol{d}]$
			\vspace*{\fill}
		\end{minipage}
		\begin{minipage}{0.8\textwidth}
			\begin{subfigure}{0.23\textwidth}
				\caption*{\textbf{Estimate}}
				\includegraphics[width=\linewidth]{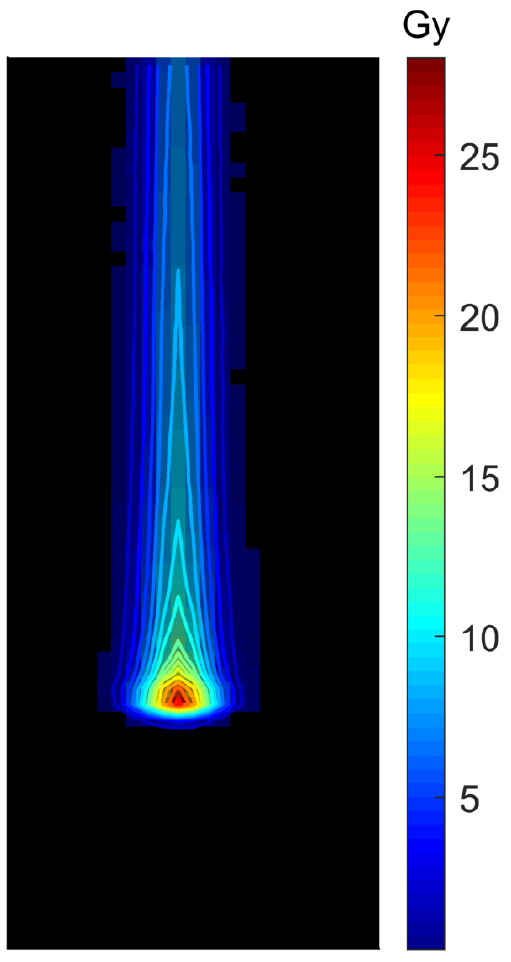}
			\end{subfigure}\hfil 
			\begin{subfigure}{0.23\textwidth}
				\caption*{\textbf{Reference}}
				\includegraphics[width=\linewidth]{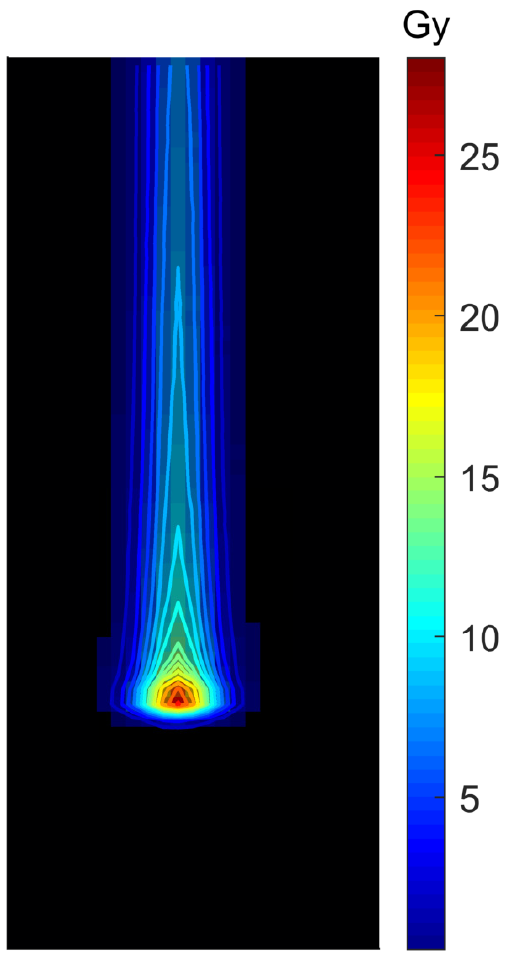}
			\end{subfigure}\hfil 
			\begin{subfigure}{0.24\textwidth}
				\caption*{\textbf{Difference}}
				\includegraphics[width=\linewidth]{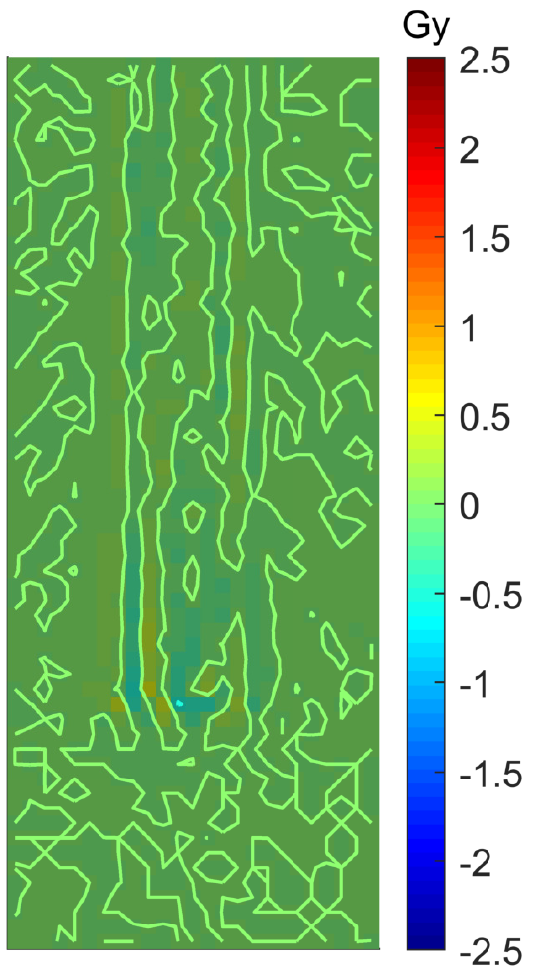}
			\end{subfigure}
			\begin{subfigure}{0.23\textwidth}
			\caption*{\textbf{$\gamma$-map}}
			\includegraphics[width=\linewidth]{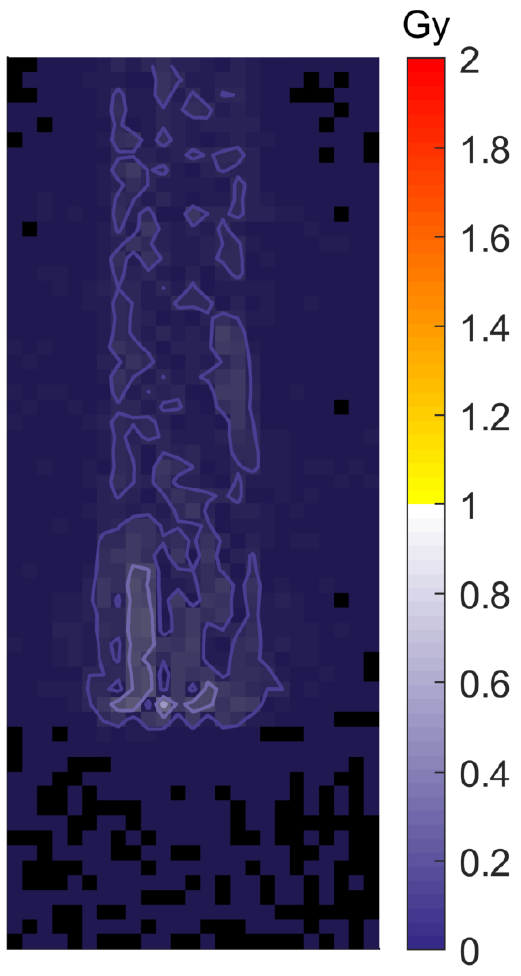}
			\end{subfigure}
		\end{minipage}\hfil
		
		\begin{minipage}{0.05\textwidth}
			\vspace*{\fill}
			$\boldsymbol{\sigma}$
			\vspace*{\fill}
		\end{minipage}
		\begin{minipage}{0.8\textwidth}
			\begin{subfigure}{0.23\textwidth}
				\includegraphics[width=\linewidth]{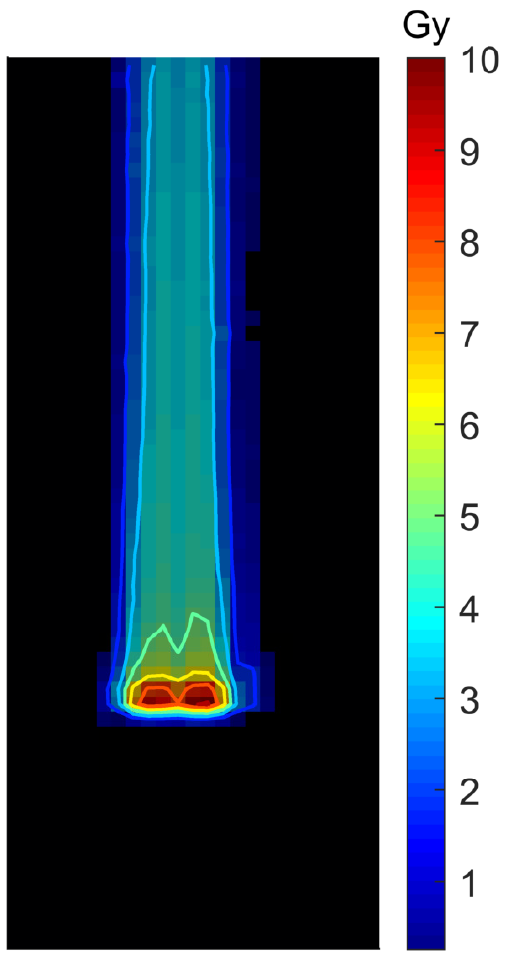}
			\end{subfigure}\hfil 
			\begin{subfigure}{0.23\textwidth}
				\includegraphics[width=\linewidth]{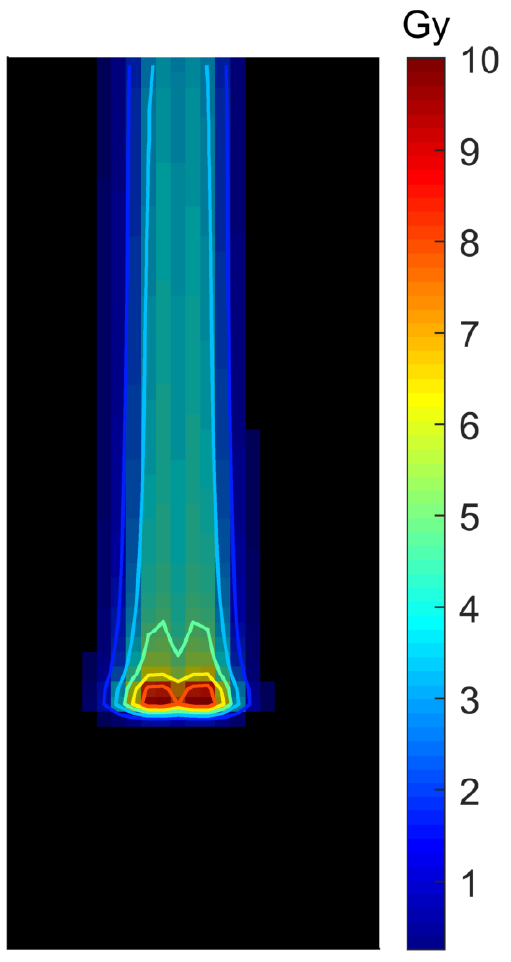}
			\end{subfigure}\hfil 
			\begin{subfigure}{0.24\textwidth}
				\includegraphics[width=\linewidth]{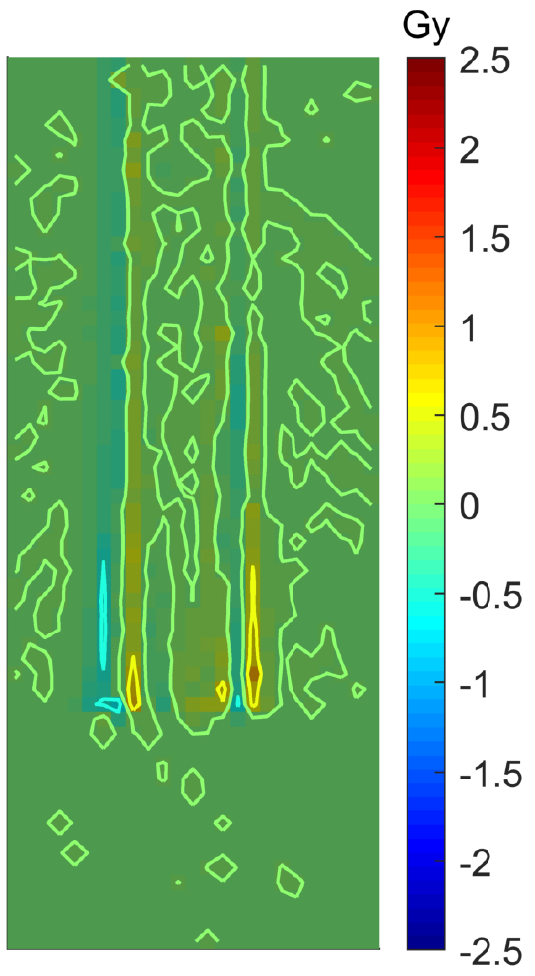}
			\end{subfigure}
			\begin{subfigure}{0.23\textwidth}
			\includegraphics[width=\linewidth]{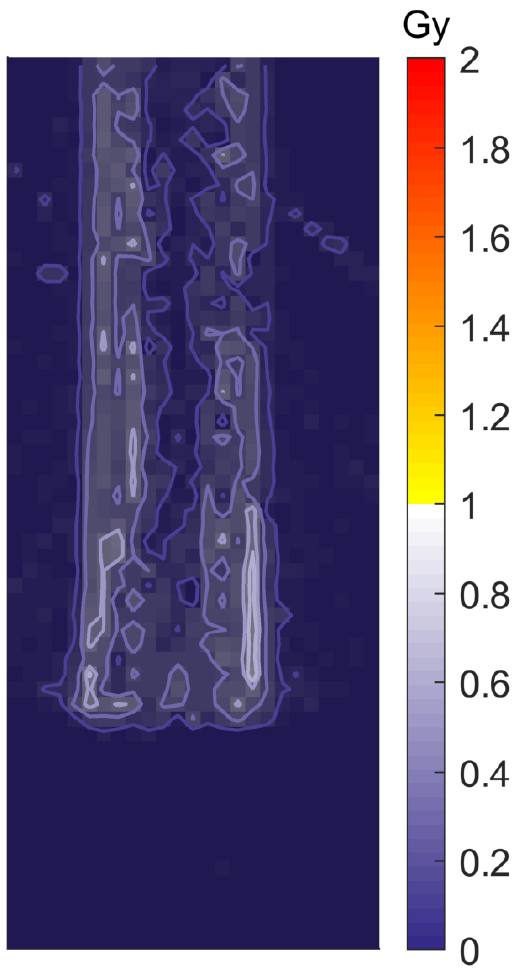}
		   \end{subfigure}
		\end{minipage}\hfil
		\caption{Expected dose and standard deviation w.r.t. set-up uncertainties with $\SI{3}{\milli\meter}$ standard deviation for one pencil beam in a water box. The left column shows the estimate computed with the proposed (re-)weighting approach, the middle column a reference computed with Monte Carlo and the right column a plot of the $\gamma_{\SI{3}{\mm}/\SI{3}{\percent}}$-indices.}
		\label{fig:WB_Setup}
	\end{figure}

\begin{figure}[H]
		\centering 	
		\begin{minipage}{0.05\textwidth}
			\vspace*{\fill}
			$E[\boldsymbol{d}]$
			\vspace*{\fill}
		\end{minipage}
		\begin{minipage}{0.9\textwidth}
			\begin{subfigure}{0.23\textwidth}
				\caption*{\textbf{Estimate}}
				\includegraphics[width=\linewidth]{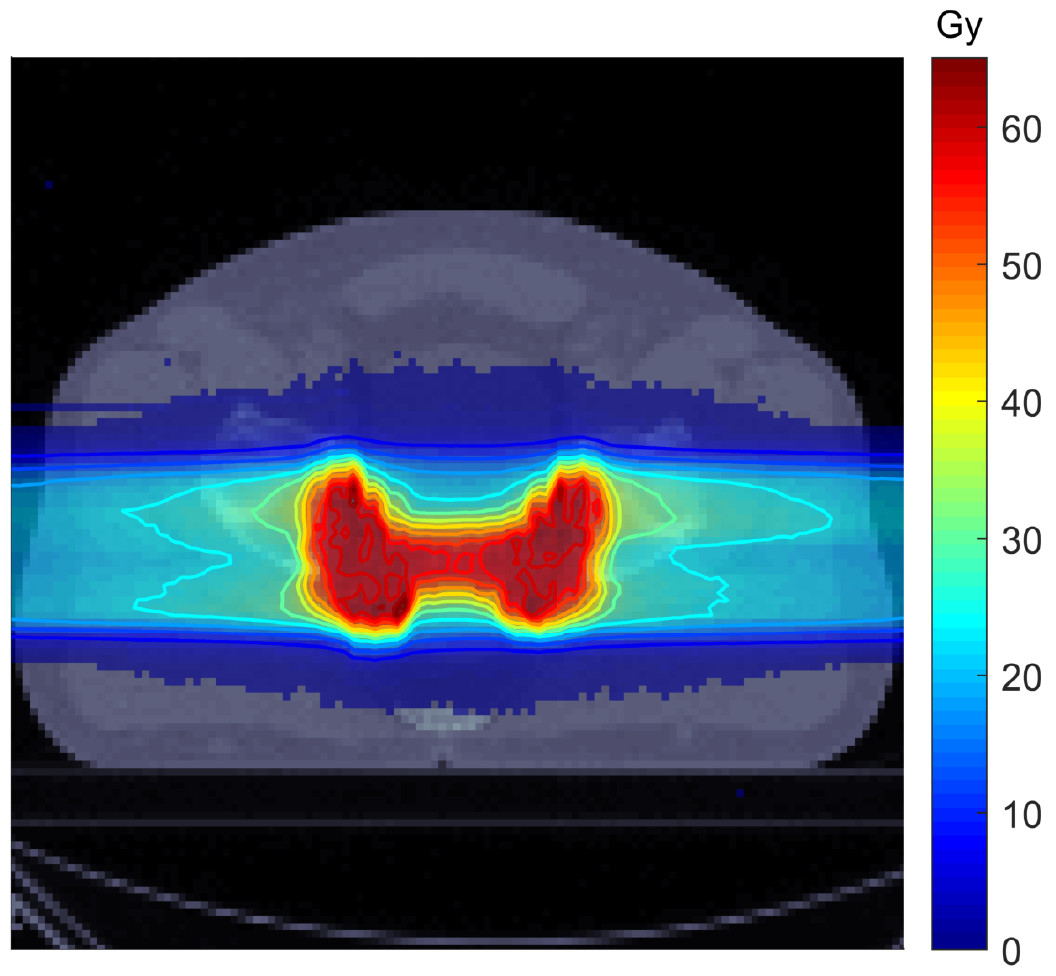}
				
			\end{subfigure}\hfil 
			\begin{subfigure}{0.23\textwidth}
				\caption*{\textbf{Reference}}
				\includegraphics[width=\linewidth]{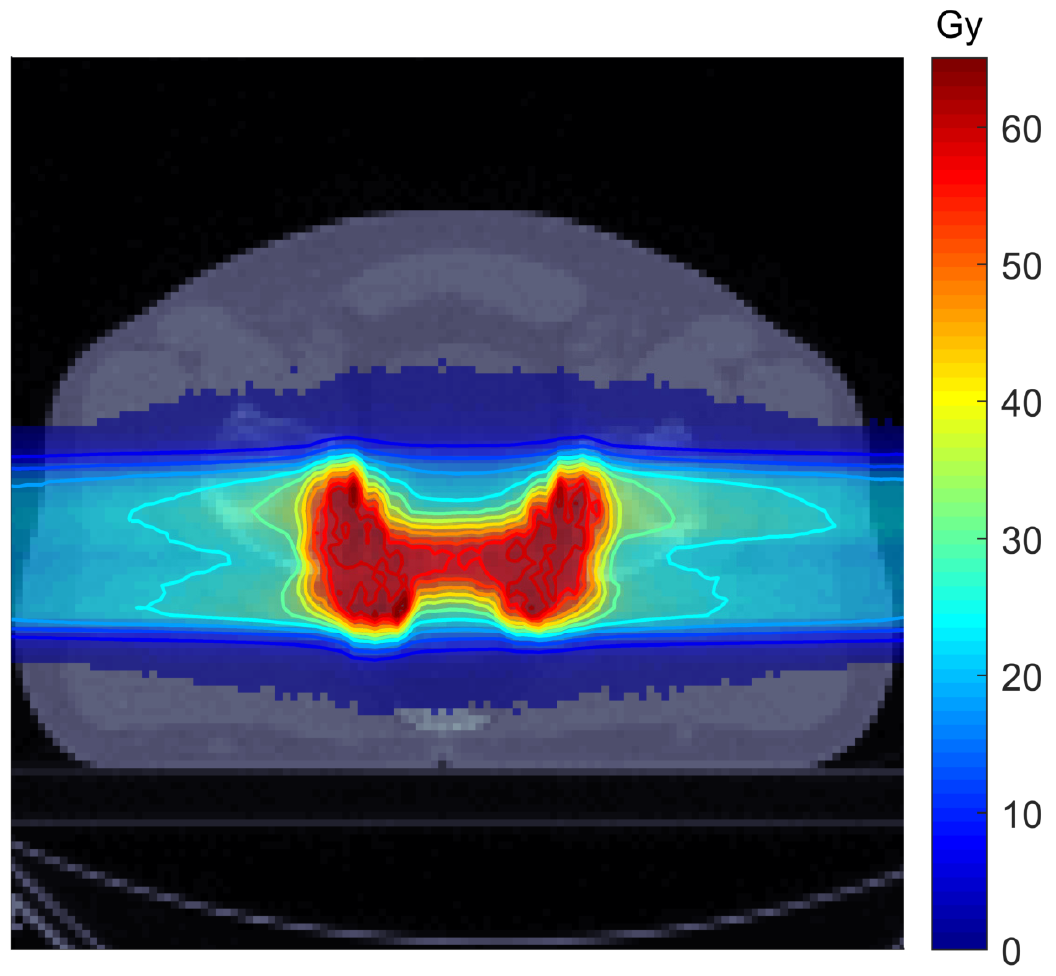}
				
			\end{subfigure}\hfil 
			\begin{subfigure}{0.235\textwidth}
				\caption*{\textbf{$\gamma$-map}}
				\includegraphics[width=\linewidth]{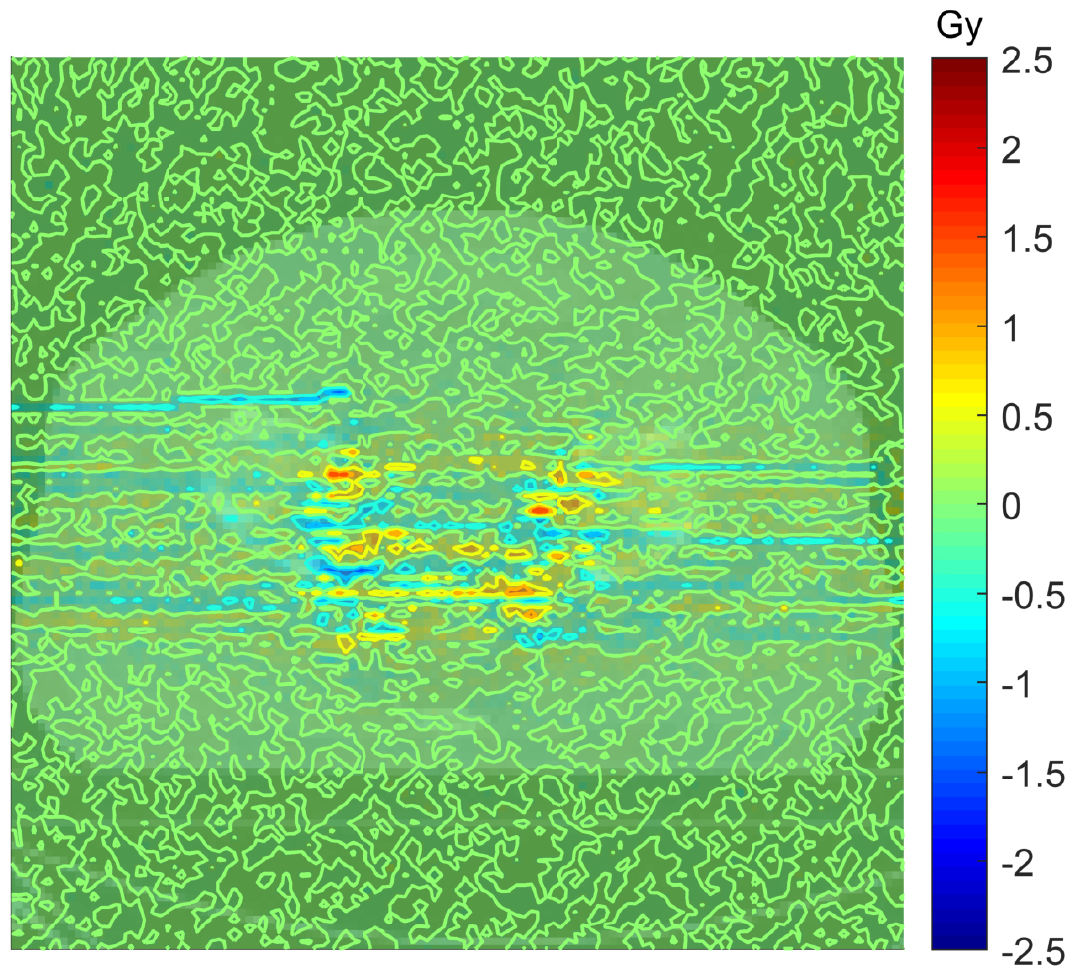}
			\end{subfigure}\hfil 
			\begin{subfigure}{0.23\textwidth}
			\caption*{\textbf{Difference}}
			\includegraphics[width=\linewidth]{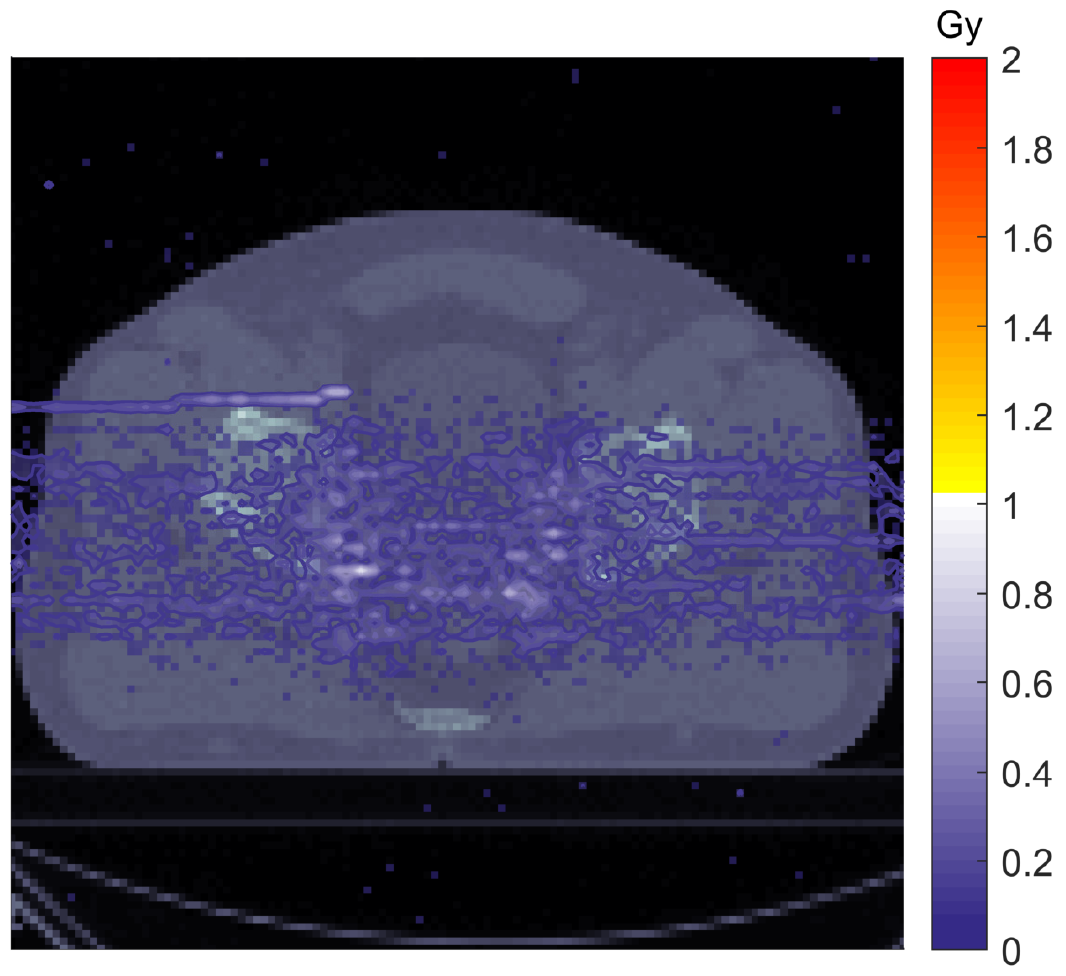}
			\end{subfigure}
		\end{minipage}\hfil
		
		\begin{minipage}{0.05\textwidth}
			\vspace*{\fill}
			$\boldsymbol{\sigma}$
			\vspace*{\fill}
		\end{minipage}
		\begin{minipage}{0.9\textwidth}
			\begin{subfigure}{0.23\textwidth}
				\includegraphics[width=\linewidth]{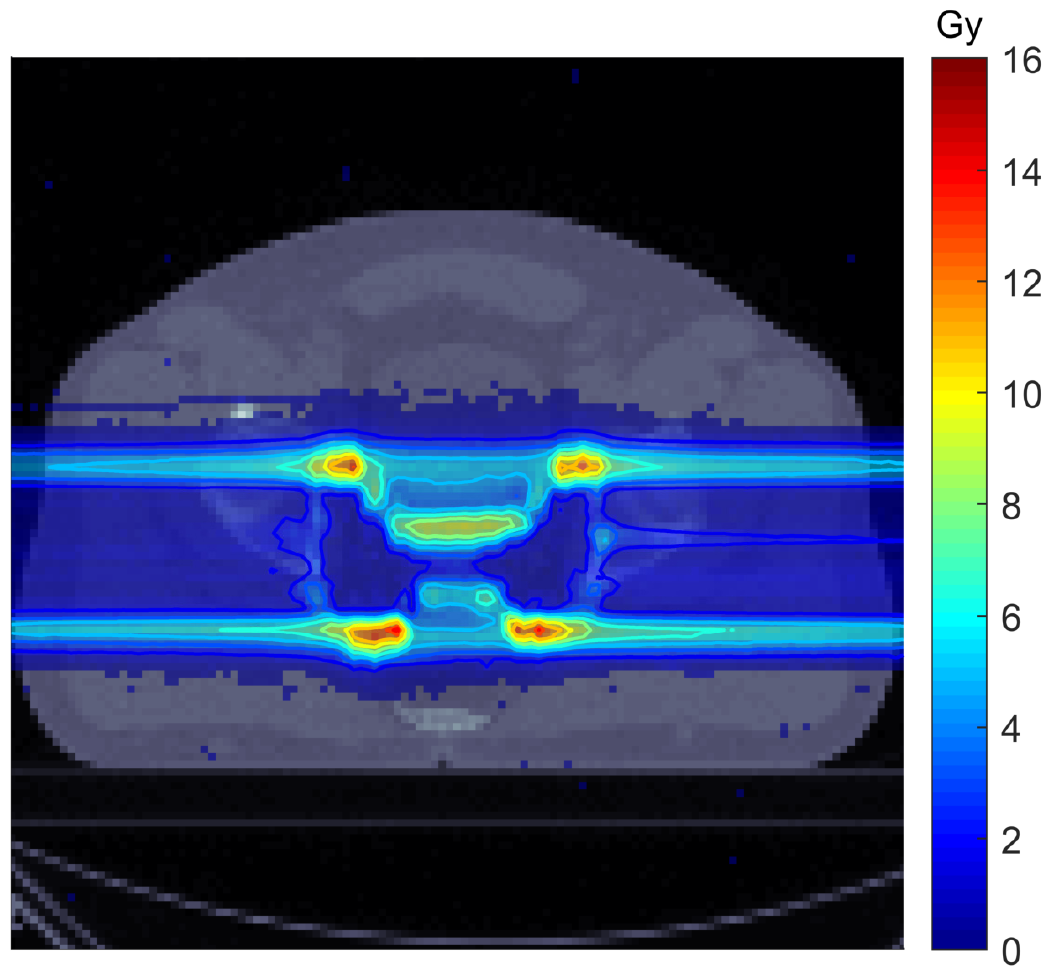}	
			\end{subfigure}\hfil 
			\begin{subfigure}{0.23\textwidth}
				\includegraphics[width=\linewidth]{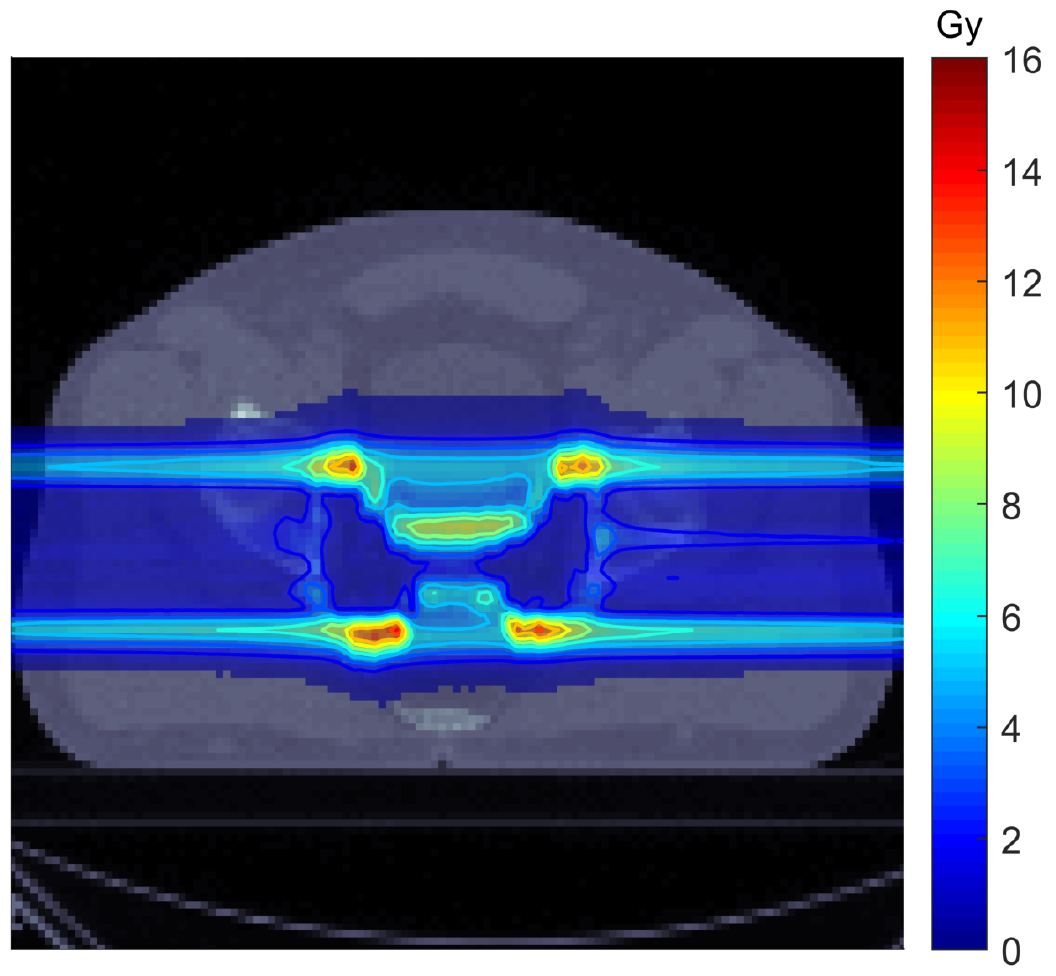}
			\end{subfigure}\hfil 
			\begin{subfigure}{0.235\textwidth}
			\includegraphics[width=\linewidth]{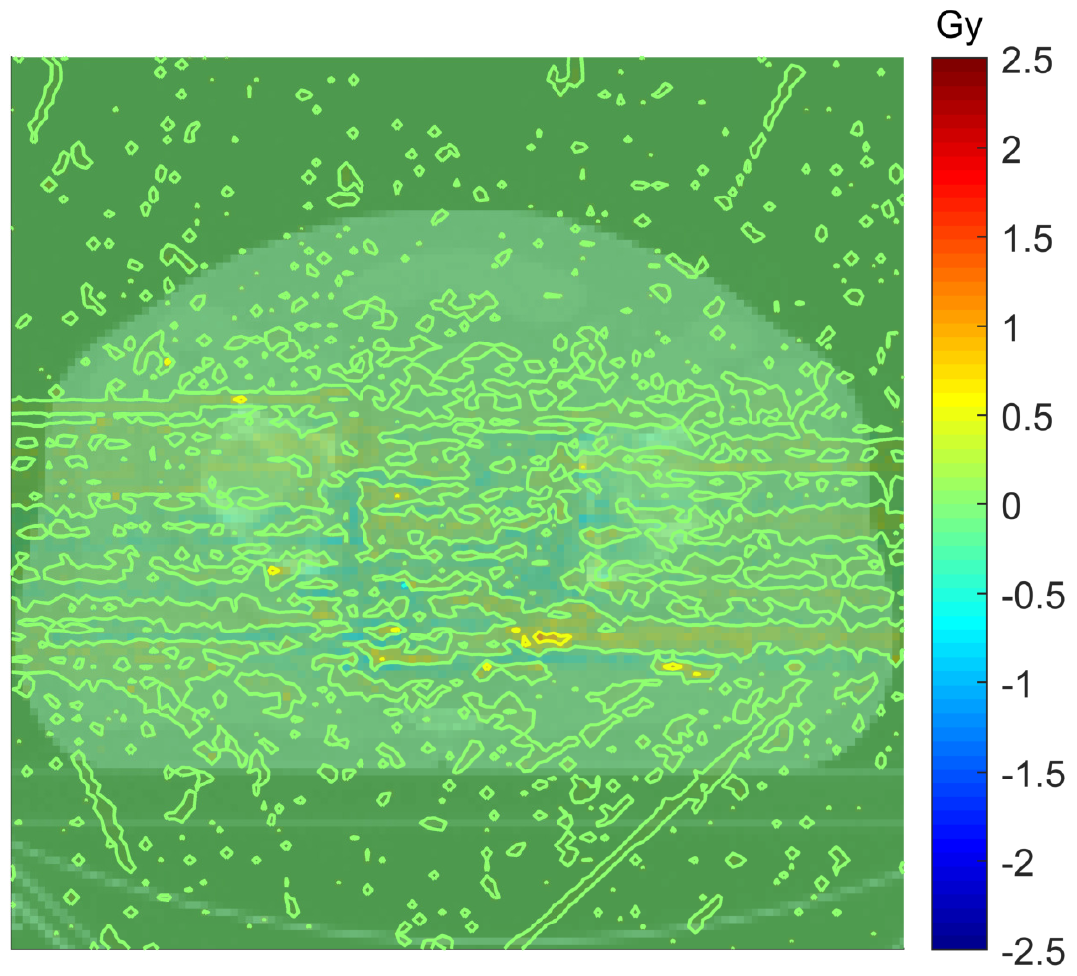}
			\end{subfigure}\hfil 
			\begin{subfigure}{0.23\textwidth}
				\includegraphics[width=\linewidth]{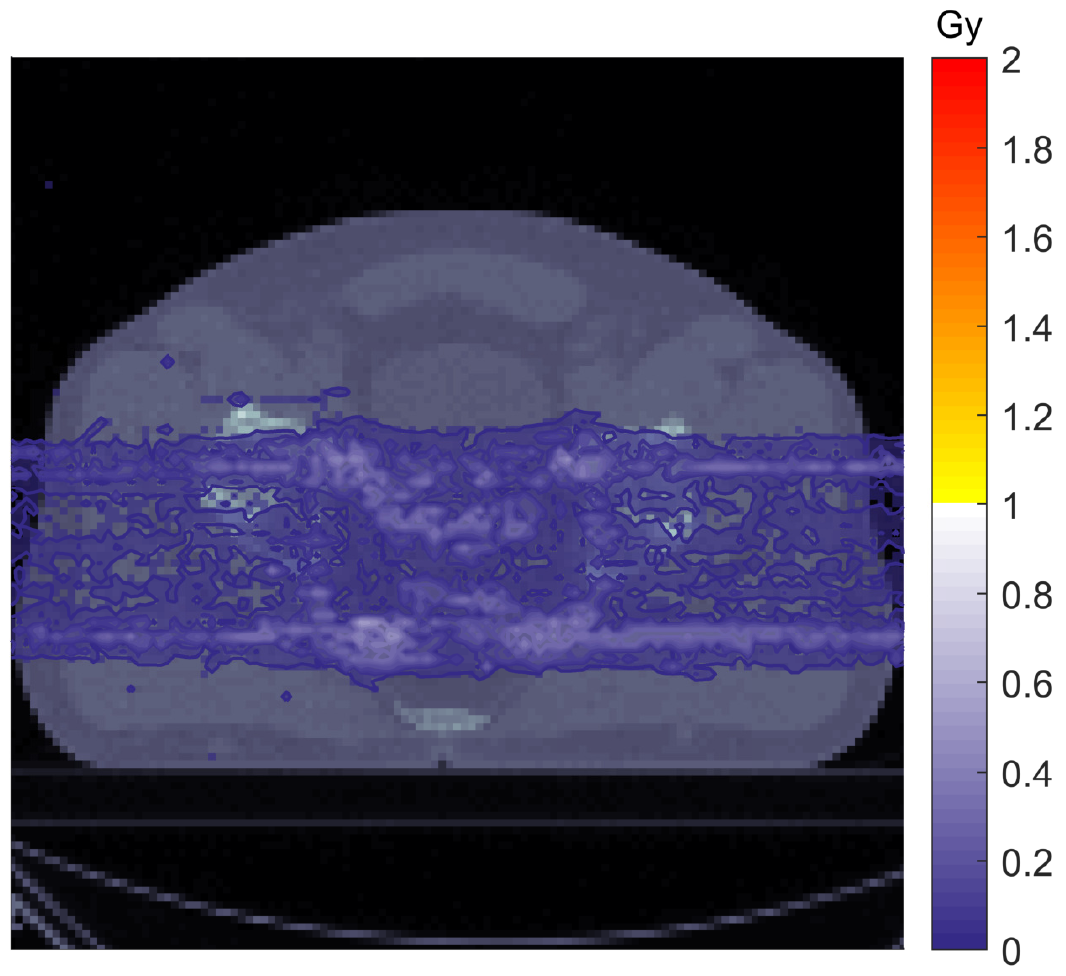}
			\end{subfigure}
		\end{minipage}\hfil
		\caption{Expected dose and standard deviation w.r.t. set-up uncertainties with $\SI{3}{\mm}$ standard deviation for two beams (gantry angles 90° and 270°, couch angle 0°) in a prostate patient. The left column shows the estimate computed with the proposed (re-)weighting approach, the middle column a reference computed with Monte Carlo and the right column the results of a $\gamma_{\SI{3}{\mm}/\SI{3}{\percent}}$-analysis.}
		\label{fig:prostate_Setup}
	\end{figure}

	\Cref{fig:WB_Setup} presents the expected dose and standard deviation for a single pencil beam in a water phantom. In order to enable a direct comparison to a more complex application case, we also include a recomputation of the respective quantities for the prostate patient investigated in \cite{stammerEfficientUncertaintyQuantification2021} with fully correlated pencil beams, i.e., one global error, in \cref{fig:prostate_Setup}. The importance distribution was chosen to be the nominal distribution $S_0$. For this reason, the nominal dose estimate is just a regular Monte Carlo estimate and is therefore omitted in the following figures.

		The results using the importance (re-)weighting estimate are in good agreement with the reference, with a $\gamma_{\SI{3}{\mm}/\SI{3}{\percent}}$-passrate of $\SI{100}{\percent}$ for the expected value and over $\SI{99.9}{\percent}$ with respect to the standard deviation in both waterbox and prostate patient.
	
	\subsection{Other correlation models}
	
\begin{figure}[H]
		\centering 
		\begin{subfigure}{0.3\textwidth}
			\caption*{\textbf{(a) Full correlation \hspace*{2mm}}}
			\includegraphics[width=\linewidth]{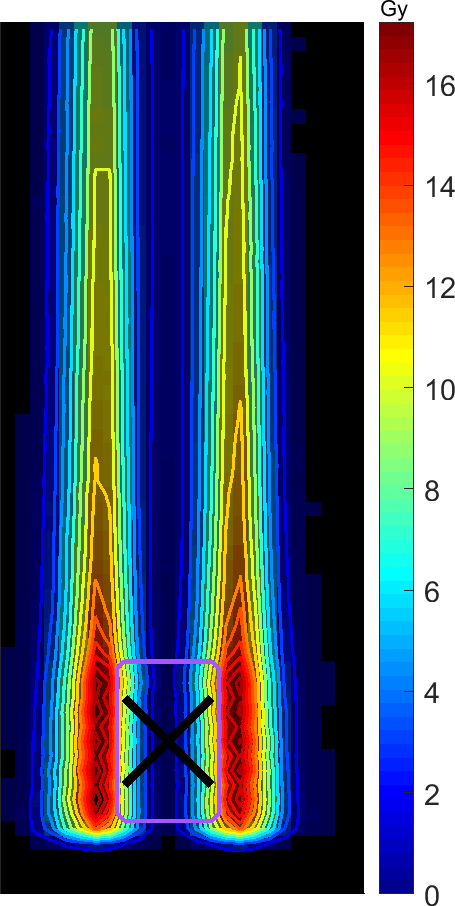}
		\end{subfigure}\hfil 
		\begin{subfigure}{0.3\textwidth}
			\caption*{\textbf{(b) AR(1) \hspace*{5mm}}}
			\includegraphics[width=\linewidth]{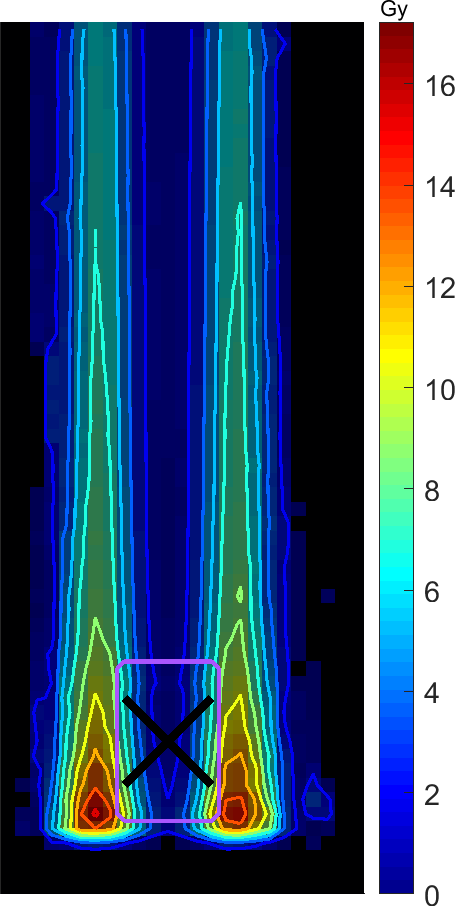}
		\end{subfigure}\hfil 
		\begin{subfigure}{0.3\textwidth}
			\caption*{\textbf{(c) Local periodic \hspace*{4mm}}}
			\includegraphics[width=\linewidth]{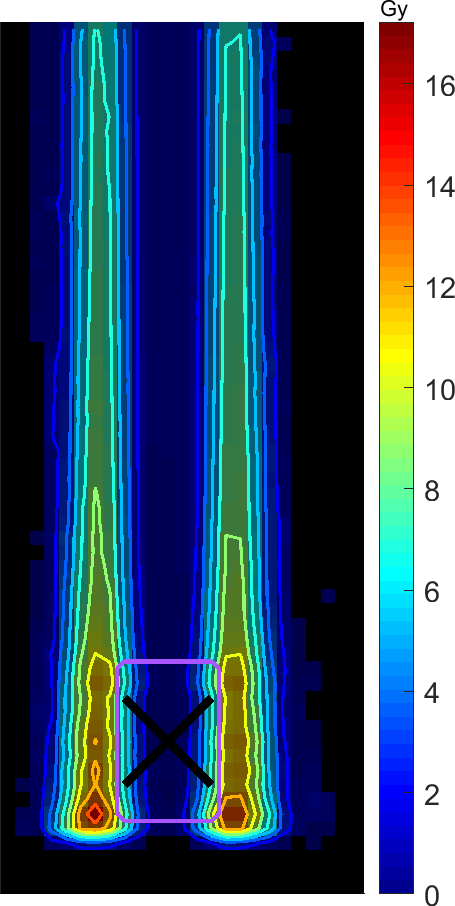}
		\end{subfigure}\hfil 
		\caption{Standard deviation of dose from a beam made up of 175 pencil beams in a 3D waterbox for different error models: (a) one global error (b) AR(1) model for movement between energy levels (c) seasonal AR(1)  model for movement between energy levels}
		\label{fig:WaterBox_Correlation}
	\end{figure}

	In \cref{sec:modelingUncertainties}, we derived correlation matrices and error distributions for different uncertainty models. Our implementation was validated against Monte Carlo results for individual realizations of shifts over time. For a single ray with $6$ energy levels in a water box, shifted by $\pm \SI{2}{mm}$ every $\SI{10}{ms}$, the dose results in each time step reached $\SI{100}{\percent}$ agreement in a $\gamma$-analysis with tolerances of $\SI{2}{\mm}/\SI{2}{\percent}$ in dose and location, respectively. \Cref{fig:WaterBox_Correlation} compares the dose standard deviation for three such models in a water phantom with a beam consisting of several superimposed pencil beams of seven different energy groups. The results illustrate that already in this simple case the uncertainty assumption has a significant impact on both the amount and shape of the dose standard deviation. However, due to the low amount of energy levels and homogeneous tissue, the characteristics of the different input models are not fully visible.
	
\begin{figure}[H]
	\centering 
	\begin{subfigure}{0.32\textwidth}
		\caption*{\textbf{(a) Full correlation \hspace*{2mm}}}
		\includegraphics[width=\linewidth]{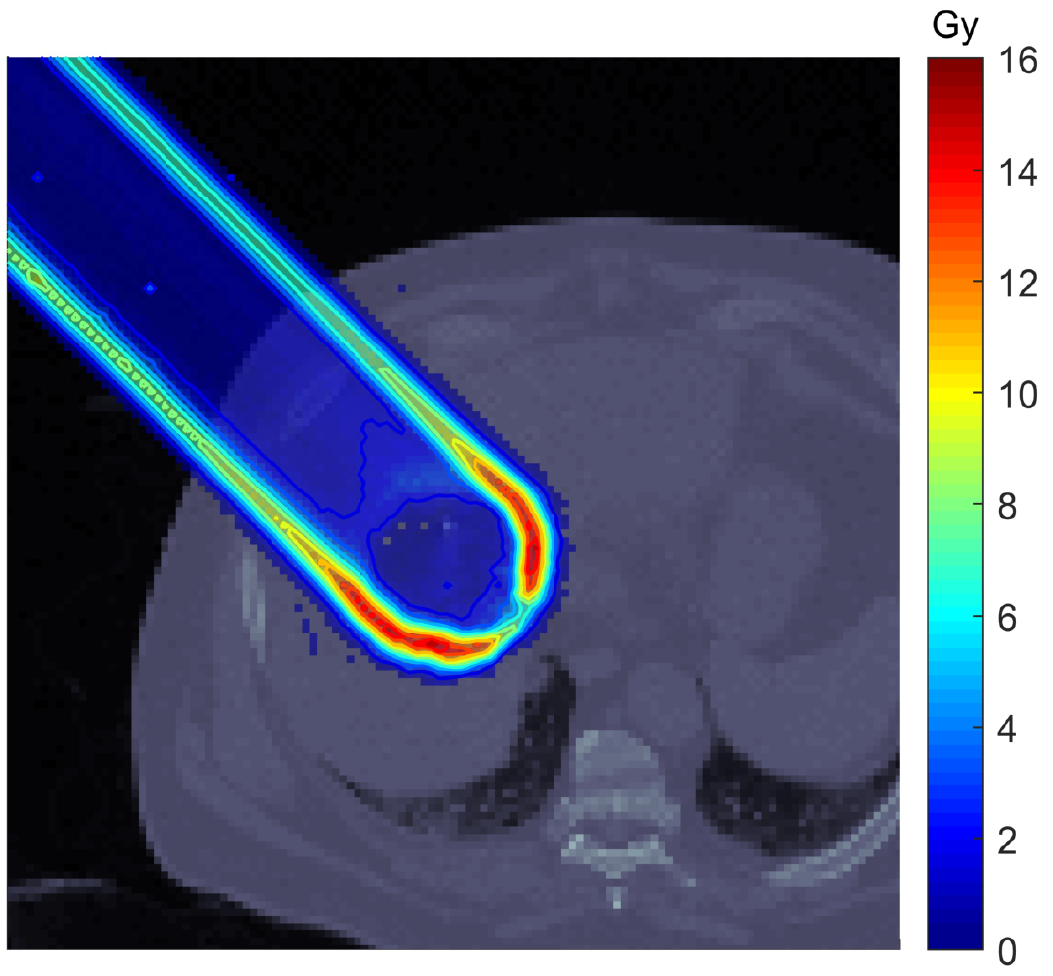}
	\end{subfigure}\hfil 
	\begin{subfigure}{0.32\textwidth}
		\caption*{\textbf{(b) AR(1) \hspace*{4mm}}}
		\includegraphics[width=\linewidth]{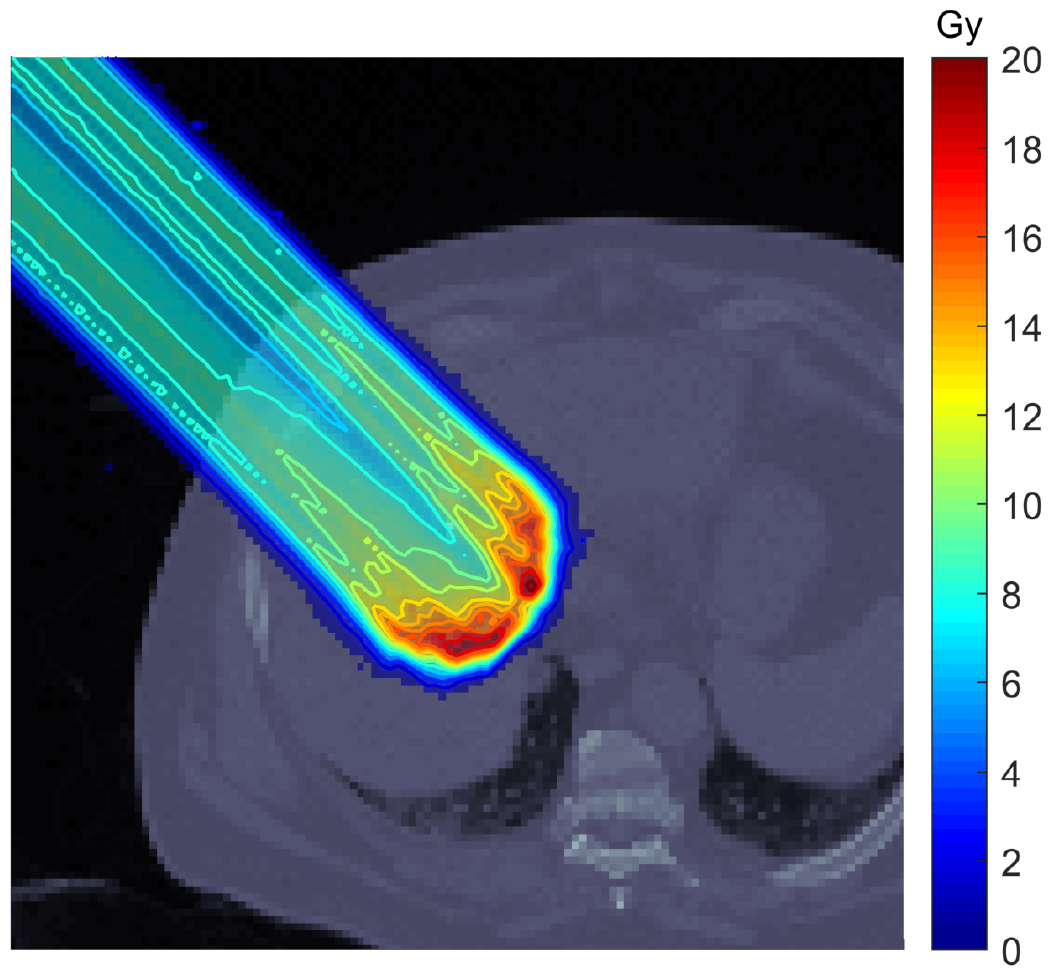}
	\end{subfigure}\hfil 
	\begin{subfigure}{0.32\textwidth}
	\caption*{\textbf{(c) Local periodic \hspace*{4mm}}}
	\includegraphics[width=\linewidth]{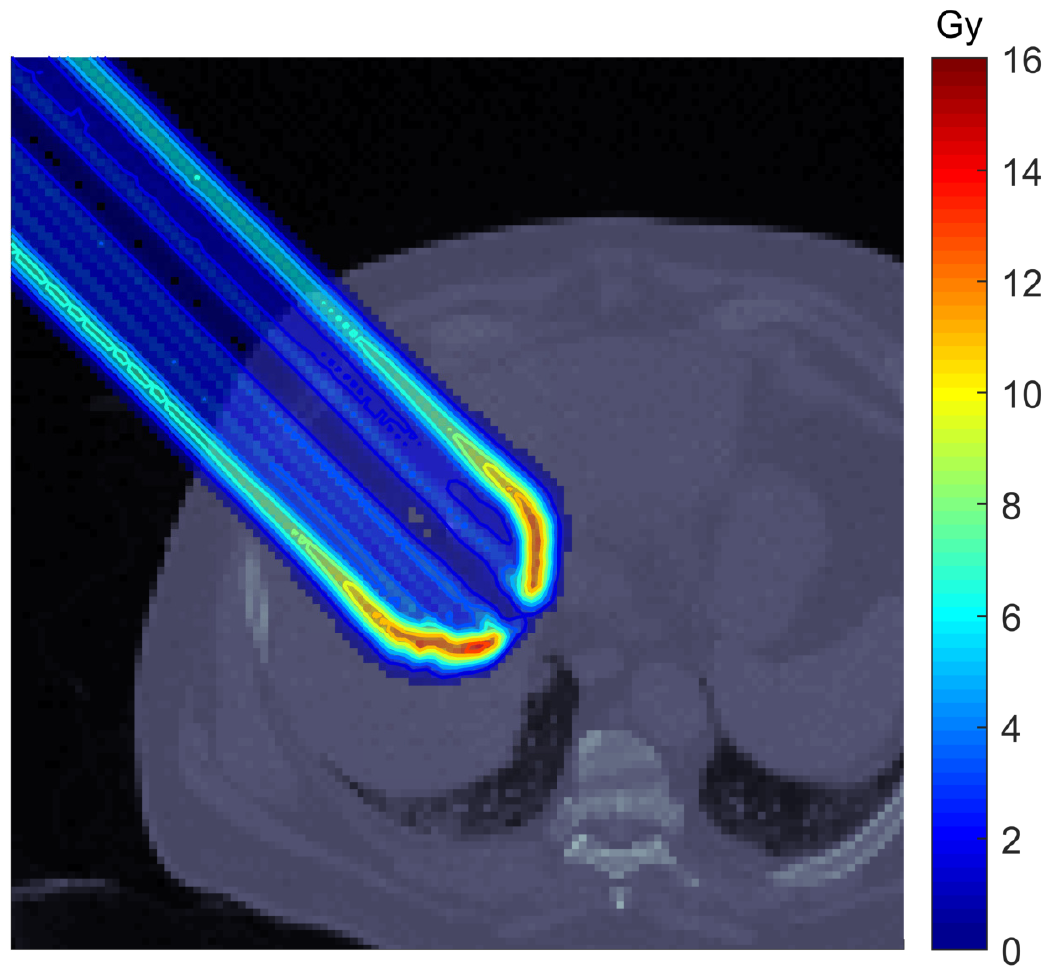}
	\end{subfigure}\hfil 
	\caption{Standard deviation of dose from a beam with irradiation angle (gantry angle $\ang{315}$, couch angle $\ang{0}$) made up of 1378 pencil beams in a liver patient for different error models: (a) one global error (b) seasonal AR(1) model for movement between energy levels}
	\label{fig:Liver_Correlation}
	\end{figure}

	This becomes apparent in \cref{fig:Liver_Correlation}, which compares the standard deviation for the three error models in a realistic liver patient, with 21 energy levels and 166 rays. While the region of high standard deviation is still at the gradients at the edge of the beam in the fully correlated and periodic model, several local peaks can additionally be observed in the latter. In this case, considering the accumulated dose over all time steps/energy levels results in a lower magnitude of standard deviation, due to averaging of the errors. In the AR(1) model, which mimics a random movement process over time, the standard deviation is less concentrated to the beam edges. Here, pencil beams are frequently shifted against instead with each other, due to the larger random component of the error in each time step/energy level.

	\subsection{Practical investigation of mathematical properties}
	\subsubsection{Bias}
\begin{figure}[H]
	\centering
		\begin{subfigure}{0.45\textwidth}
			\includegraphics[width=\linewidth]{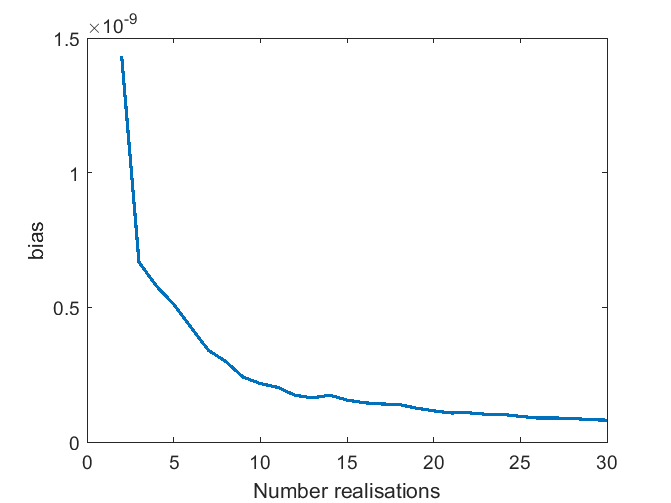}
			\caption*{(a)}
		\end{subfigure}
		\begin{subfigure}{0.45\textwidth}
			\includegraphics[width=\linewidth]{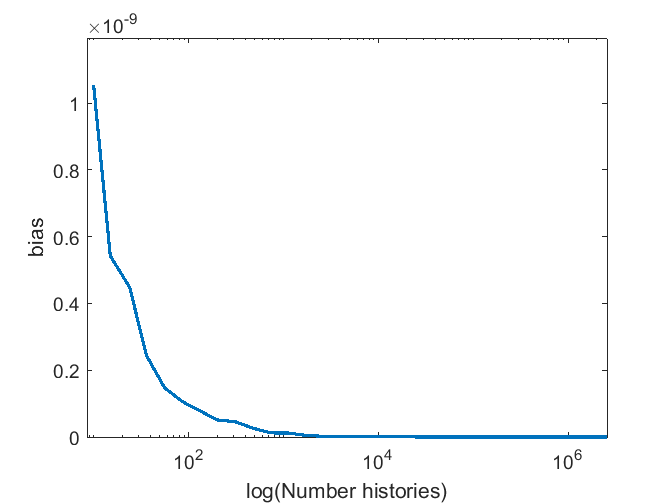}
			\caption*{(b)}
		\end{subfigure}
		\caption{Convergence of the variance estimator's bias in (a) number of error realizations used to compute the bias and (b) the number of particle histories used for the dose reconstructions for a single pencil beam in a water box.}
		\label{BiasConvergence}
	\end{figure}

	The above findings indicate that the bias in the dose variance estimate, introduced in proposition \ref{prop:2}, does not significantly affect the accuracy of the results, as the agreement with the reference solutions is comparable for the unbiased expected value estimator and the biased variance estimator. Further, \cref{BiasConvergence} shows that the bias reduces for an increasing number of particle histories.
	
	\subsubsection{Standard error}
	\label{sec:numError}
	So far, the dose for all estimates was reconstructed from samples of the distribution $S_0$. However, we have seen in \cref{TheoreticalError}, that the standard error of the estimates depends on the relation of distribution from which the original sample was drawn to the target distribution. When computing the dose standard deviation, each error scenario corresponds to a Gaussian input distribution with a shifted mean value (see \cref{sec:modelingUncertainties}). \Cref{AccuracyperDistPlots} shows that the accuracy of the estimates decreases with growing distance of the mean of the target distributions from that of the sampling distribution. This development can be observed not only for the theoretical standard error, but also for the mean square error and $\gamma$-passrates. This indicates that the more extreme realizations of the uncertain parameters introduce relatively high errors to the overall standard deviation estimate. We want to see, whether the use of a better suited sampling distribution, such as a wider Gaussian or a Gaussian mixture of some of the target distributions, can improve the accuracy of the estimate.
	
	For this, we sample input parameters once from the joint distribution $\mathcal{S}$ and once from a mixture of the narrower Gaussian $S_0$, shifted by $\pm \sigma$ and $\pm 2\sigma$ in all coordinate directions. In order to better detect differences in the results, we compare the estimates constructed from these samples with the results from $S_0$ using the stricter criteria {\SI{2}{\mm}/\SI{2}{\percent} for the $\gamma$-analysis.
	
	\begin{figure}[H]
		\begin{subfigure}{0.33\textwidth}
			\includegraphics[width=\linewidth]{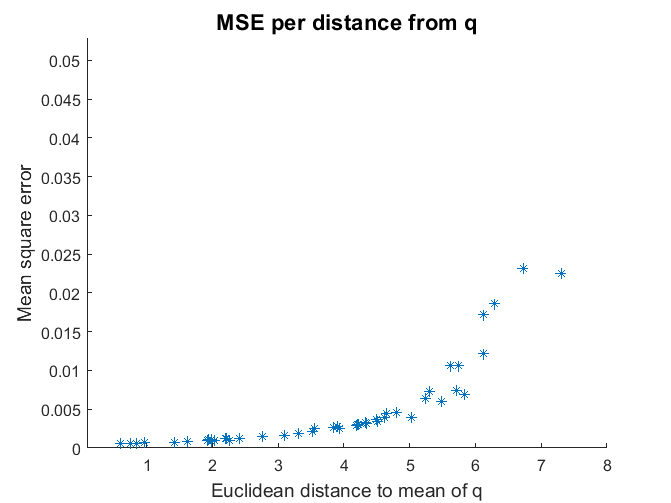}
			\caption*{(a)}
		\end{subfigure}
		\begin{subfigure}{0.33\textwidth}
			\includegraphics[width=\linewidth]{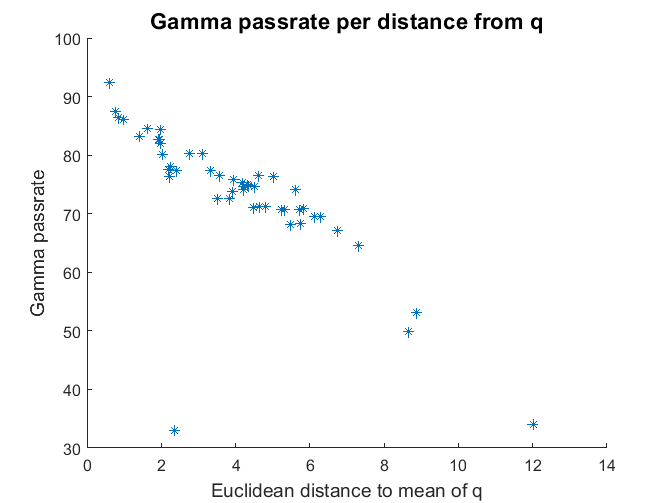}
			\caption*{(b)}
		\end{subfigure}
		\begin{subfigure}{0.33\textwidth}
			\includegraphics[width=\linewidth]{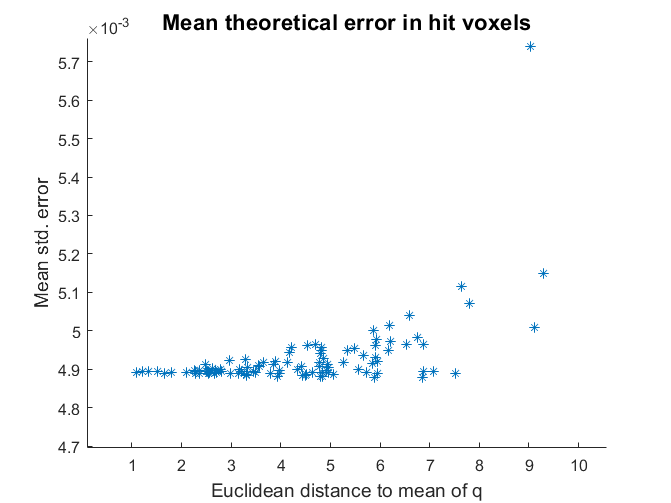}
			\caption*{(c)}
		\end{subfigure}
		\caption{Accuracy of dose estimates for increasing distance between importance distribution $q$ and target distribution $\Phi_i$. (a) Measured by mean square error (MSE), (b) by $\gamma$-passrate and (c) the mean of the theoretical error (see \cref{TheoreticalError}) in voxels with non-zero dose.}
		\label{AccuracyperDistPlots}
	\end{figure}
	
	\Cref{fig:WaterBox_DifferentDist} illustrates that the $\gamma$-passrate increases when using both the wider Gaussian and the Gaussian mixture. Thus, especially when one has an uncertainty model involving errors with a high variance or if the expected value and standard deviation of the dose are more important than the nominal dose, it can be worthwhile to not use the regular parameter distribution to generate the initial sample. For medical purposes, the high accuracy of the nominal dose is however an integral part of the quality assurance of a method and could often be prioritized over the quantification of uncertainties.

	\begin{figure}[H]
		\centering 
		\begin{subfigure}{0.3\textwidth}
			\caption*{\textbf{(a) $S_0$}, $\gamma_{\SI{2}{\mm}/\SI{2}{\percent}}$-passrate = \SI{94.64}{\percent}}
			\includegraphics[width=\linewidth]{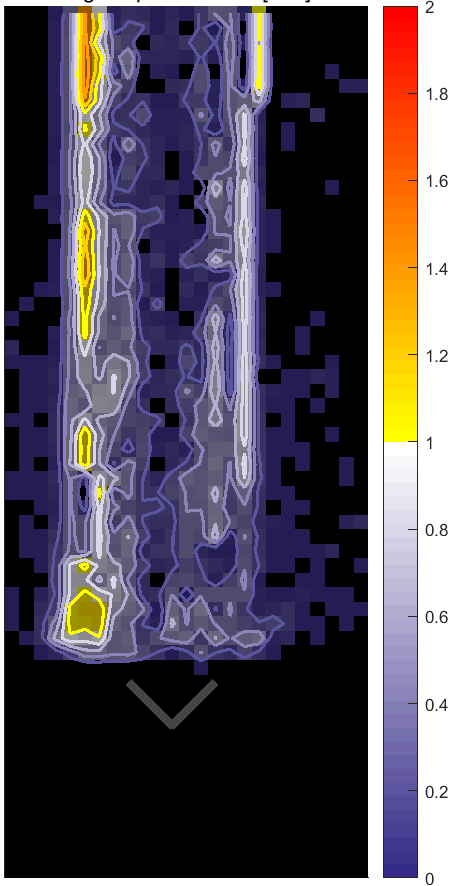}
		\end{subfigure}\hfil 
		\begin{subfigure}{0.3\textwidth}
			\caption*{\textbf{(b) $\mathcal{S}$},  $\gamma_{\SI{2}{\mm}/\SI{2}{\percent}}$-passrate = \SI{99.86}{\percent}}
			\includegraphics[width=\linewidth]{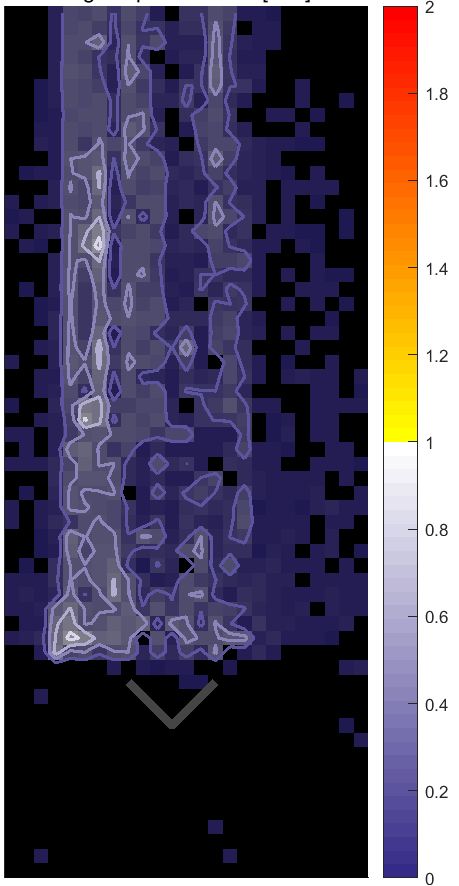}
		\end{subfigure}\hfil 
		\begin{subfigure}{0.3\textwidth}
			\caption*{\textbf{(c)} Gaussian mixture,  $\gamma_{\SI{2}{\mm}/\SI{2}{\percent}}$-passrate = \SI{99.98}{\percent}}
			\includegraphics[width=\linewidth]{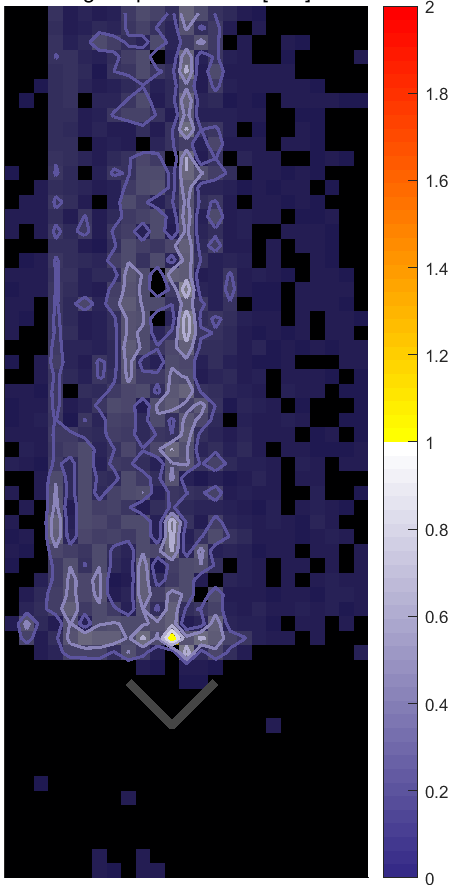}
		\end{subfigure}\hfil 
		\caption{Estimates of the dose standard deviation in a water box using different sampling distributions: (a) the nominal distribution $S_0$, (b) the distribution $\mathcal{S}$ corresponding to the expected value and (c) a Gaussian mixture distribution}
		\label{fig:WaterBox_DifferentDist}
	\end{figure}
	
	\subsubsection{Variance upper bound}
	Lastly, in \cref{sec:varianceBound} an upper bound for the dose variance estimate was derived, which is significantly less computationally expensive than the variance estimate itself. \Cref{fig:WaterBox_UpperBound} shows a comparison of this upper bound estimate to the dose variance. While the variance values are expectedly higher in the upper bound estimate, both exhibit a similar structure with two distinct variance peaks at positions with high dose gradients. This indicates a potential use for optimization purposes. However, further analyses in more complex test cases, in particular such with several irradiataion angles and in heterogeneous materials, are necessary.
	\begin{figure}[H]
		\centering 
		\begin{subfigure}{0.3\textwidth}
			\caption*{\hspace{-0.5cm}\textbf{Upper bound of dose variance}}
			\includegraphics[width=\linewidth]{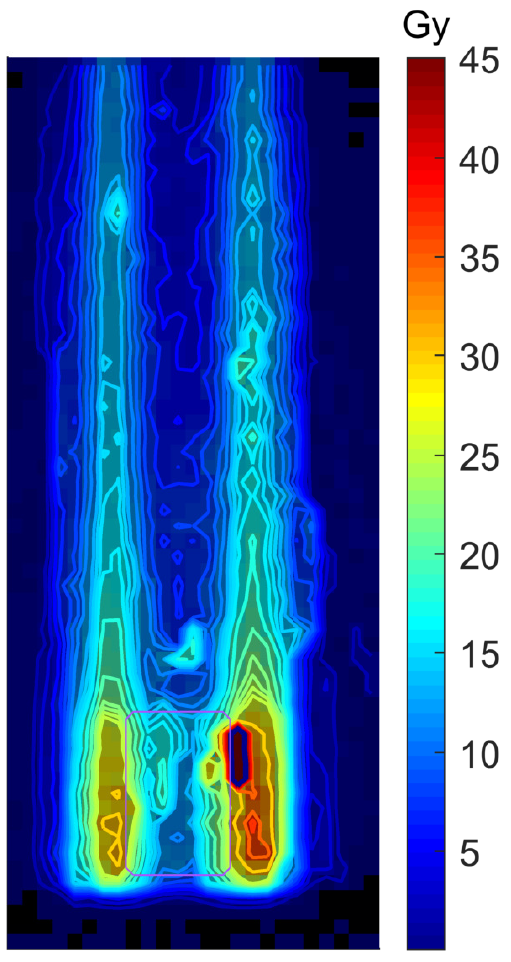}
		\end{subfigure}\hfil 
		\begin{subfigure}{0.3\textwidth}
			\caption*{\hspace{-0.75cm}\textbf{Dose variance}}
			\includegraphics[width=\linewidth]{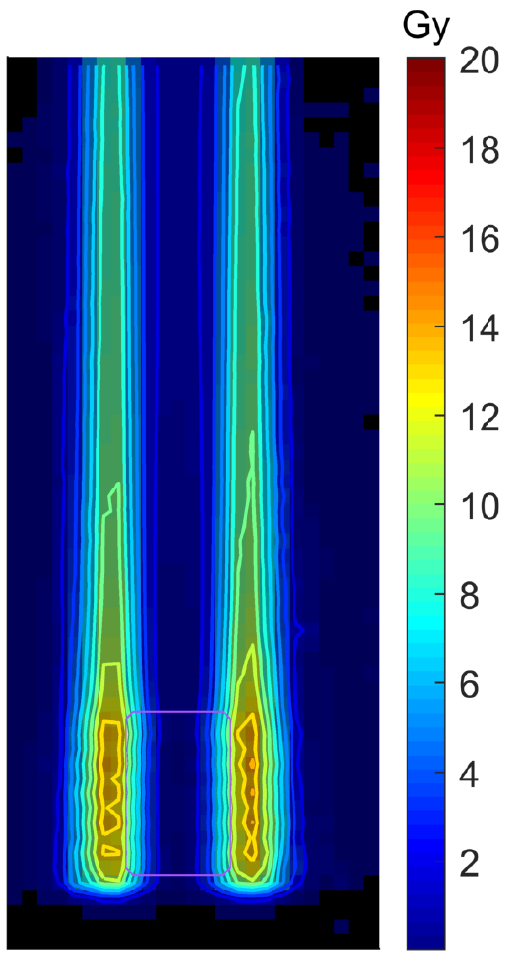}
		\end{subfigure}\hfil 
		\begin{subfigure}{0.3\textwidth}
			\caption*{\hspace{-1cm}\textbf{Difference}}
			\includegraphics[width=\linewidth]{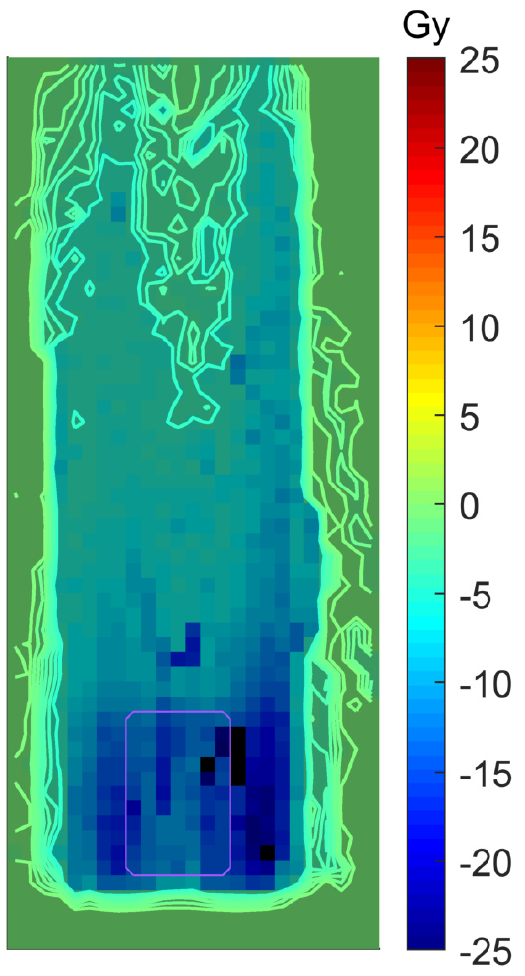}
		\end{subfigure}\hfil 
		\caption{Comparison of the upper bound to the dose variance estimate for a beam made up of 175 pencil beams in a 3D water box.}
		\label{fig:WaterBox_UpperBound}
	\end{figure}
	
	\subsection{Convergence}

	The proposed approach mimics a sampling-based uncertainty quantification method. The solution of the modified problem for each error realization using the black box solver is replaced by (re-)weighting the dose corresponding to particle trajectories from a previous simulation. Thus, the convergence of the method is equivalent to that of the underlying UQ method, in our case randomized quasi-Monte Carlo. \Cref{convergencePlots} illustrates this for the dose standard deviation of the single beam in a water box. While the convergence of reference and estimate is very similar per computed realization, the speed-up is achieved by the quicker computation of each iteration, that is, only performing MC scoring operations compared to physical simulations. Therefore, the amount of time which can be saved also depends on the specific patient case, treatment plan and software, which determine how much overhead is produced for example by the initialization and simulation of physical processes.
	
	\begin{figure}[H]
		\begin{subfigure}{0.45\textwidth}
			\includegraphics[width=\linewidth]{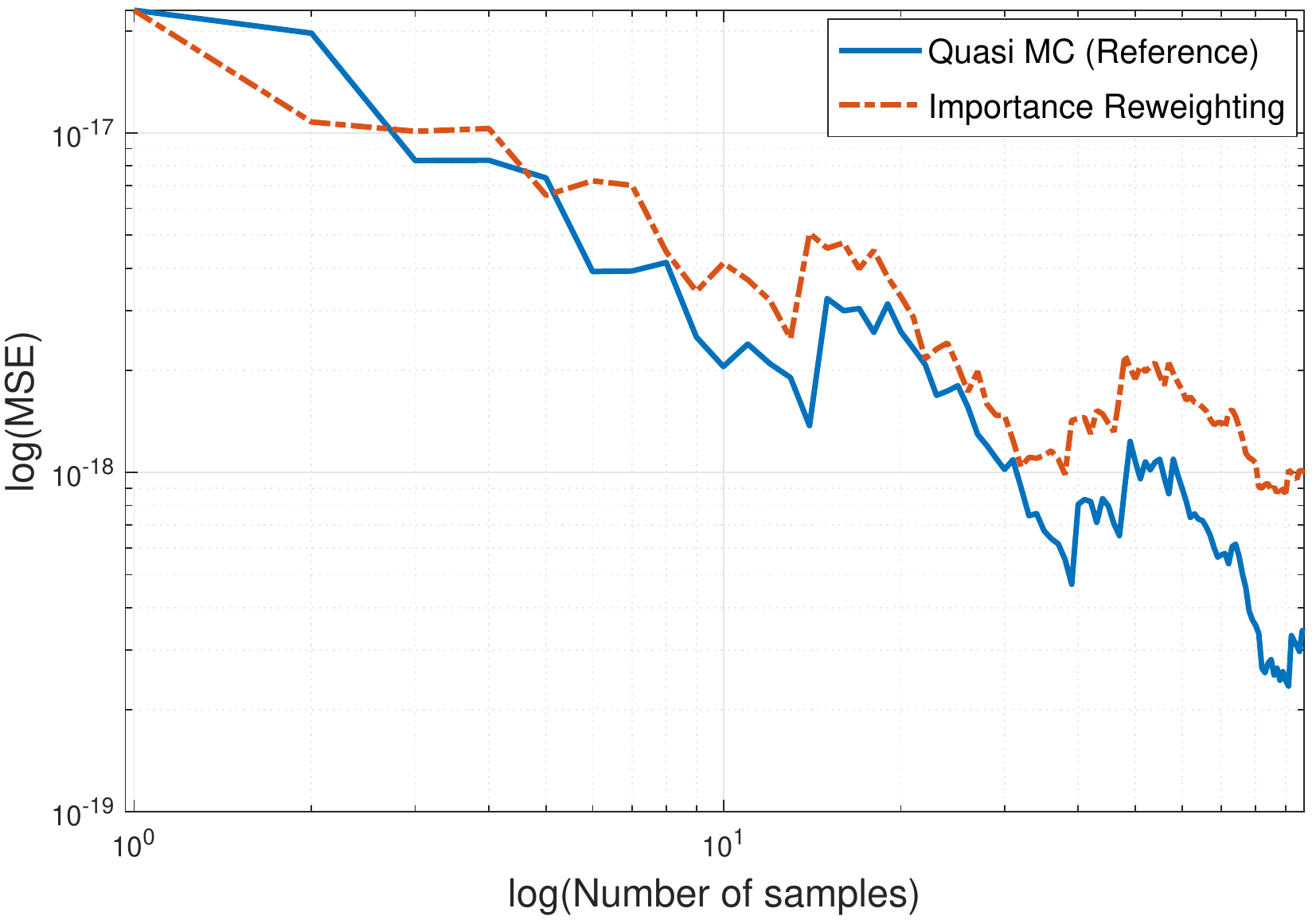}
			\caption*{(a)}
		\end{subfigure}
		\begin{subfigure}{0.45\textwidth}
			\includegraphics[width=\linewidth]{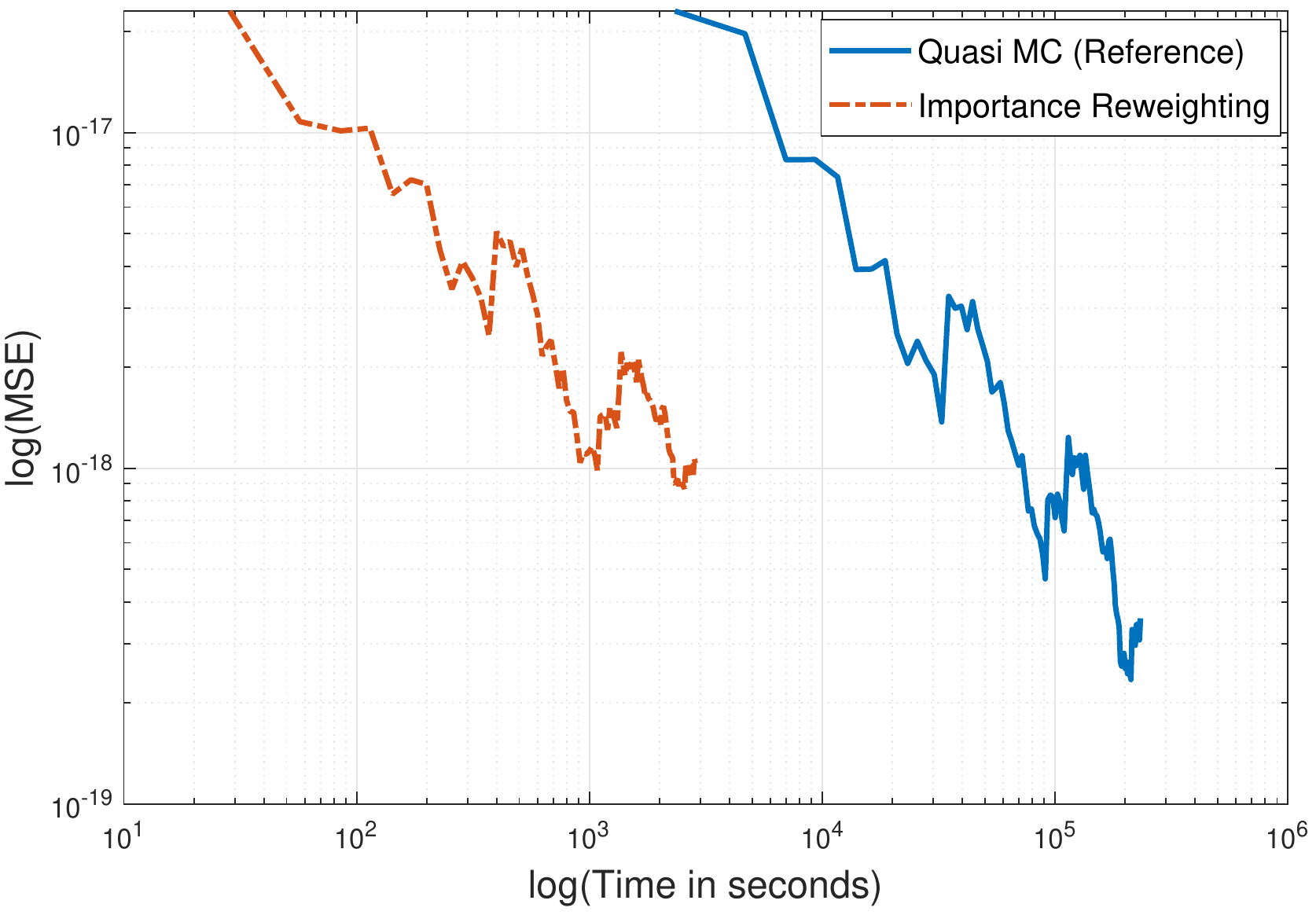}
			\caption*{(b)}
		\end{subfigure}
		\caption{Mean squared error of different methods compared (a) per iteration and (b) per time.}
		\label{convergencePlots}
	\end{figure}

	For run-time comparisons, reference computations and the (re-)weighting approach were run on the same virtual machine\footnote{Virtual machine including 64 CPUs with 1.995 GHz and 200GB RAM} and timed during the computation of the dose standard deviation using the global error model. As \cref{table:CPUtimes} shows, we observe reduced CPU times by a factor of approximately 183 and 80 for the two treatment plans using the water phantom and factors of 32 and 23 for the liver and prostate patient. Note, that the large difference in initialization times between the IMPT plans and the plan involving just one beam is due to the size of the sparse matrix containing the doses for all particle histories. For the larger plans, this exceeded the available RAM such that the use of less efficient tall arrays was necessary. Note, that these conclusions refer solely to a post-processing implementation of the method. When implemented as on-the-fly scoring no storage of histories is required, instead, one sparse dose cube per considered scenario has to be stored. Also, the possibility to use existing efficient structures of the Monte Carlo code might enable performance improvements compared to the post-processing routine.
	
	\begin{table}[htb!]
		\centering
		\caption{CPU time comparison for the reference vs. (re-)weighting approach applied to different test cases and computed on the same machine. All values are given in seconds. Note that the times for 100 realizations include the initialization times, while the time for a single realization only refers to the dose computation time.}
		\begin{tabular}{l l l l}
			\hline
			& & reference& (re-)weighting \\
			\hline
			\multirow{3}{*}{Water phantom (1 beam)}& Initialization & 2.27& 1.34 \\
			& One realization &72.21  & 0.38\\
			& 100 realizations & 7221.64& 39.34\\
			&&&\\[-2ex]
			\multirow{3}{*}{Water phantom (IMPT)}& Initialization & 2.35& 61.53 \\
			& One realization &2331.30  & 28.51\\
			& 100 realizations & 233126.93& 2912.53\\
			 &&&\\[-2ex]
            \multirow{3}{*}{Liver}		  & Initialization &2.44 &2038.75 \\
      								  & One realization &39066.44  &1198.74 \\
        							  & 100 realizations &3906650.90 &121912.75 \\
        	
			&&&\\[-2ex]
			\multirow{3}{*}{ Prostate}	  & Initialization & 4.26& 4867.75\\
			& One realization & 58762.40 & 2479.07\\
			& 100 realizations & 5876253.86&252774.75 \\
			\hline
		\end{tabular}
		\label{table:CPUtimes}
	\end{table}
	
	\section{Discussion}
	\label{Discussion}

	In this paper we demonstrate how importance sampling can be adapted to lower the computational costs of sampling-based uncertainty quantification in Monte Carlo dose calculations using the example of proton irradiation. We derive an unbiased estimator and the standard error for the expected value, as well as an estimator, bias and upper bound for the standard deviation. We apply the proposed method to Gaussian set-up uncertainties with $\SI{3}{\mm}$ standard deviation in a homogeneous water phantom, a prostate and a liver patient.  For a simple, global uncertainty model we observe high agreements of at least $\SI{99.9}{\percent}$ for both test cases (\cref{fig:WB_Setup,fig:prostate_Setup}). We show that these results can be further improved by adapting the distribution function used for the initial Monte Carlo simulation to the quantity which is of highest interest. Here the user can choose between the exactness of the nominal dose estimate and that of the expected dose or standard deviation. 
	
	The sampling of error realizations is completely independent from the solution of the transport problem, which allows us to incorporate complex correlation models including movement patterns and time dependence. This is a clear advantage compared to classic sampling based methods, which would require elaborate set up and a large number of realizations to account for the degrees of freedom in a multivariate model. Therefore, more complex correlation models are currently often neglected in uncertainty studies \cite{bangertAnalyticalProbabilisticModeling2013,pflugfelderWorstCaseOptimization2008,liuRobustOptimizationIntensity2012}, although it is apparent in  \cref{fig:WaterBox_Correlation,fig:Liver_Correlation}, that different uncertainty models can have a significant impact on the dose standard deviation, already in a homogeneous material. As an extension to this work, the time-dependent error modeling introduced in \cref{sec:modelingUncertainties} might be extended to simulate spatial motion for 4D treatment planning or even deformations in geometry components \cite[][]{bangertAnalyticalProbabilisticModeling2013}. The derived correlation models are of interest in their own right and can be transferred to any dose calculation and uncertainty quantification framework which supports multivariate error correlations \cite[e.g.][]{bangertAnalyticalProbabilisticModeling2013}. Also, the resulting uncertainty estimates may be used for the purpose of robust optimization during treatment planning.
	
	While an extension to other types of uncertainties in the input parameters is straightforward, the method is limited to errors which have a non-trivial density function. Further work could explore the adaptation and application of the model to uncertainties beyond errors in the patient set-up.
	
	The main computational advantage of the proposed method is that it minimizes the overhead associated with complex simulations. Therefore we cannot claim a faster convergence rate and the speed-up is dependent on the specific application and implementation. A combination with more sophisticated scenario-based UQ methods, such as quadratures or a non-intrusive polynomial chaos expansion, could however lead to further run-time improvements.
	
	Further, we argue that computational performance can be improved through a more efficient implementation. In particular, importance weighting can be implemented on-the-fly using multiple sub-scoring routines, where each routine accumulates the delivered dose for one scenario (see \cref{alg:IRW_postprocess_vs_onthefly}). Here, performance gains should be achievable compared to the current post-processing implementation, since the existing efficient structures of the Monte Carlo code could be directly used.  However, in this case the number of error scenarios has to be predetermined, causing a loss in flexibility, and is also limited with regard to memory constraints. 	
	
	Lastly, the proposed (re-)weighting approach is not limited to proton therapy. An application to uncertainties in radiation therapy using e.g. carbon ions or photons is feasible and the method can in principal be relevant to any case of input uncertainties in complex Monte Carlo simulations.

\section*{Acknowledgments}
The present contribution is supported by the Helmholtz Association under the joint research school HIDSS4Health — Helmholtz Information and Data Science School for Health.

	\bibliographystyle{abbrv}

\begin{thebibliography}{10}

\bibitem{bangertAnalyticalProbabilisticModeling2013}
M.~Bangert, P.~Hennig, and U.~Oelfke.
\newblock Analytical probabilistic modeling for radiation therapy treatment
  planning.
\newblock {\em Physics in Medicine and Biology}, 58(16):5401--5419, July 2013.

\bibitem{beckmanMonteCarloEstimation1987}
R.~J. Beckman and M.~D. McKay.
\newblock Monte carlo estimation under different distributions using the same
  simulation.
\newblock {\em Technometrics}, 29(2):153--160, 1987.

\bibitem{bedfordCalculationAbsorbedDose2019}
J.~L. Bedford.
\newblock Calculation of absorbed dose in radiotherapy by solution of the
  linear {{Boltzmann}} transport equations.
\newblock {\em Physics in Medicine \& Biology}, 64(2):02TR01, Jan. 2019.

\bibitem{benhamouFewPropertiesSample2018}
E.~Benhamou.
\newblock {A few properties of sample variance}.
\newblock working paper or preprint, Feb. 2019.

\bibitem{caflischMonteCarloQuasiMonte1998}
R.~E. Caflisch.
\newblock {Monte Carlo and quasi-Monte Carlo methods}.
\newblock {\em Acta Numerica}, 7:1--49, Jan. 1998.

\bibitem{casiraghiAdvantagesLimitationsWorst2013}
M.~Casiraghi, F.~Albertini, and A.~Lomax.
\newblock Advantages and limitations of the 'worst case scenario' approach in
  {{IMPT}} treatment planning.
\newblock {\em Physics in medicine and biology}, 58:1323--1339, Feb. 2013.

\bibitem{chuRobustOptimizationIntensity2005}
M.~Chu, Y.~Zinchenko, S.~G. Henderson, and M.~B. Sharpe.
\newblock Robust optimization for intensity modulated radiation therapy
  treatment planning under uncertainty.
\newblock {\em Physics in Medicine and Biology}, 50(23):5463--5477, Dec. 2005.

\bibitem{craftSharedDataIntensity2014}
D.~Craft, M.~Bangert, T.~Long, D.~Papp, and J.~Unkelbach.
\newblock Shared data for intensity modulated radiation therapy ({{IMRT}})
  optimization research: The {{CORT}} dataset.
\newblock {\em GigaScience}, 3(1), Dec. 2014.

\bibitem{davisonNeutronTransportTheory1957}
B.~Davison and J.~B. Sykes.
\newblock {\em Neutron Transport Theory}.
\newblock {Clarendon Press/Oxford University Press}, {Oxford/London}, 1957.

\bibitem{duderstadtTransportTheory1979}
J.~J. Duderstadt and W.~R. Martin.
\newblock {\em Transport Theory}.
\newblock {John Wiley \& Sons}, {New York}, 1979.

\bibitem{durichenMultitaskGaussianProcesses2015}
R.~D{\"u}richen, M.~A.~F. Pimentel, L.~Clifton, A.~Schweikard, and D.~A.
  Clifton.
\newblock Multitask {{Gaussian}} processes for multivariate physiological
  time-series analysis.
\newblock {\em IEEE transactions on bio-medical engineering}, 62(1):314--322,
  Jan. 2015.

\bibitem{duvenaudAutomaticModelConstruction2014}
D.~Duvenaud.
\newblock {\em Automatic model construction with Gaussian processes}.
\newblock PhD thesis, University of Cambridge, 2014.

\bibitem{frankApproximateModelsRadiative2007}
M.~Frank.
\newblock Approximate models for radiative transfer.
\newblock {\em Bulletin of the Institute of Mathematics. Academia Sinica. New
  Series}, 2, Jan. 2007.

\bibitem{fredrikssonScenariobasedGeneralizationRadiation2015}
A.~Fredriksson and R.~Bokrantz.
\newblock The scenario-based generalization of radiation therapy margins.
\newblock {\em Physics in Medicine and Biology}, 61, Oct. 2015.

\bibitem{fredrikssonMinimaxOptimizationHandling2011}
A.~Fredriksson, A.~Forsgren, and B.~H{\aa}rdemark.
\newblock Minimax optimization for handling range and setup uncertainties in
  proton therapy.
\newblock {\em Medical Physics}, 38(3):1672--1684, Mar. 2011.

\bibitem{giffordComparisonFiniteelementMultigroup2006}
K.~A. Gifford, J.~L. Horton, T.~A. Wareing, G.~Failla, and F.~Mourtada.
\newblock Comparison of a finite-element multigroup discrete-ordinates code
  with {{Monte Carlo}} for radiotherapy calculations.
\newblock {\em Physics in Medicine and Biology}, 51(9):2253--2265, Apr. 2006.

\bibitem{hachiyaAdaptiveImportanceSampling2009}
H.~Hachiya, T.~Akiyama, M.~Sugiayma, and J.~Peters.
\newblock Adaptive importance sampling for value function approximation in
  off-policy reinforcement learning.
\newblock {\em Neural Networks}, 22(10):1399--1410, 2009.

\bibitem{hastingsMonteCarloSampling1970}
W.~K. Hastings.
\newblock Monte {{Carlo Sampling Methods Using Markov Chains}} and {{Their
  Applications}}.
\newblock {\em Biometrika}, 57(1):97--109, 1970.

\bibitem{hesterbergWeightedAverageImportance1995}
T.~Hesterberg.
\newblock Weighted average importance sampling and defensive mixture
  distributions.
\newblock {\em Technometrics}, 37(2):185--194, 1995.

\bibitem{hommaNewMotionManagement2009}
N.~Homma, M.~Sakai, H.~Endo, M.~Mitsuya, Y.~Takai, and M.~Yoshizawa.
\newblock A new motion management method for lung tumor tracking radiation
  therapy.
\newblock {\em WSEAS Transactions on Systems}, 8:471--480, Apr. 2009.

\bibitem{huTimeSeriesAnalysis2012}
A.~Hu and S.~Qi.
\newblock Time {{Series Analysis}} of {{Interfraction Patient Setup}} in
  {{Image Guided Radiation Therapy}}.
\newblock {\em International Journal of Radiation Oncology, Biology, Physics},
  84(3):S736--S737, Nov. 2012.

\bibitem{husseinChallengesCalculationGamma2017}
M.~Hussein, C.~Clark, and A.~Nisbet.
\newblock Challenges in calculation of the gamma index in radiotherapy
  \textendash{} {{Towards}} good practice.
\newblock {\em Physica Medica}, 36:1--11, Apr. 2017.

\bibitem{jabbariReviewFastMonte2011}
K.~Jabbari.
\newblock Review of {{Fast Monte Carlo Codes}} for {{Dose Calculation}} in
  {{Radiation Therapy Treatment Planning}}.
\newblock {\em Journal of medical signals and sensors}, 1:73--86, Jan. 2011.

\bibitem{kahnRandomSamplingMonte1950}
H.~Kahn.
\newblock Random sampling ({{Monte Carlo}}) techniques in neutron attenuation
  problems--{{I}}.
\newblock {\em Nucleonics}, 6(5):27; passim, May 1950.

\bibitem{kanaiSpotScanningSystem1980}
T.~Kanai, K.~Kawachi, Y.~Kumamoto, H.~Ogawa, T.~Yamada, H.~Matsuzawa, and
  T.~Inada.
\newblock Spot scanning system for proton radiotherapy.
\newblock {\em Medical Physics}, 7(4):365--369, 1980.

\bibitem{kollig2002efficient}
T.~Kollig and A.~Keller.
\newblock Efficient multidimensional sampling.
\newblock {\em Computer Graphics Forum}, 21(3):557--563, 2002.

\bibitem{kraanDoseUncertaintiesIMPT2013}
A.~C. Kraan, S.~{van de Water}, D.~N. Teguh, A.~{Al-Mamgani}, T.~Madden, H.~M.
  Kooy, B.~J.~M. Heijmen, and M.~S. Hoogeman.
\newblock Dose uncertainties in {{IMPT}} for oropharyngeal cancer in the
  presence of anatomical, range, and setup errors.
\newblock {\em International Journal of Radiation Oncology, Biology, Physics},
  87(5):888--896, Dec. 2013.

\bibitem{laine2011stratified}
S.~Laine and T.~Karras.
\newblock {Stratified Sampling for Stochastic Transparency}.
\newblock {\em Computer Graphics Forum}, 2011.

\bibitem{lEcuyer2018randomized}
P.~L'Ecuyer.
\newblock Randomized quasi-monte carlo: An introduction for practitioners.
\newblock In A.~B. Owen and P.~W. Glynn, editors, {\em Monte Carlo and
  Quasi-Monte Carlo Methods}, pages 29--52, Cham, 2018. Springer International
  Publishing.

\bibitem{liuRobustOptimizationIntensity2012}
W.~Liu, X.~Zhang, Y.~Li, and R.~Mohan.
\newblock Robust optimization of intensity modulated proton therapy.
\newblock {\em Medical Physics}, 39(2):1079--1091, Feb. 2012.

\bibitem{lomaxIntensityModulationMethods1999}
A.~Lomax.
\newblock Intensity modulation methods for proton radiotherapy.
\newblock {\em Physics in Medicine and Biology}, 44(1):185--205, Jan. 1999.

\bibitem{lomaxIntensityModulatedProton2008}
A.~J. Lomax.
\newblock Intensity modulated proton therapy and its sensitivity to treatment
  uncertainties 1: The potential effects of calculational uncertainties.
\newblock {\em Physics in medicine and biology}, 53(4):1027--1042, Feb. 2008.

\bibitem{lomaxIntensityModulatedProton2008a}
A.~J. Lomax.
\newblock Intensity modulated proton therapy and its sensitivity to treatment
  uncertainties 2: The potential effects of inter-fraction and inter-field
  motions.
\newblock {\em Physics in medicine and biology}, 53(4):1043--1056, Feb. 2008.

\bibitem{lowTechniqueQuantitativeEvaluation1998}
D.~A. Low, W.~B. Harms, S.~Mutic, and J.~A. Purdy.
\newblock A technique for the quantitative evaluation of dose distributions.
\newblock {\em Medical Physics}, 25(5):656--661, May 1998.

\bibitem{matousek1998L2discrepancy}
J.~Matoušek.
\newblock On the l2-discrepancy for anchored boxes.
\newblock {\em Journal of Complexity}, 14(4):527--556, 1998.

\bibitem{mcgowanTreatmentPlanningOptimisation2013}
S.~E. McGowan, N.~G. Burnet, and A.~J. Lomax.
\newblock Treatment planning optimisation in proton therapy.
\newblock {\em The British Journal of Radiology}, 86(1021):20120288--20120288,
  Jan. 2013.

\bibitem{owen1995randomly}
A.~B. Owen.
\newblock Randomly permuted (t, m, s)-nets and (t, s)-sequences.
\newblock In {\em Monte Carlo and quasi-Monte Carlo methods in scientific
  computing}, pages 299--317. Springer, 1995.

\bibitem{owenMCVariance1997}
A.~B. Owen.
\newblock Monte carlo variance of scrambled net quadrature.
\newblock {\em SIAM Journal on Numerical Analysis}, 34(5):1884--1910, 1997.

\bibitem{owenMonteCarloTheory2013}
A.~B. Owen.
\newblock {\em Monte {{Carlo}} Theory, Methods and Examples}.
\newblock 2013.

\bibitem{paganettiRangeUncertaintiesProton2012}
H.~Paganetti.
\newblock Range uncertainties in proton therapy and the role of {{Monte Carlo}}
  simulations.
\newblock {\em Physics in Medicine and Biology}, 57(11):R99--R117, June 2012.

\bibitem{parkStatisticalAssessmentProton2013}
P.~C. Park, J.~P. Cheung, X.~R. Zhu, A.~K. Lee, N.~Sahoo, S.~L. Tucker, W.~Liu,
  H.~Li, R.~Mohan, L.~E. Court, and L.~Dong.
\newblock Statistical assessment of proton treatment plans under setup and
  range uncertainties.
\newblock {\em International journal of radiation oncology, biology, physics},
  86(5):1007--1013, Aug. 2013.

\bibitem{perkoFastAccurateSensitivity2016}
Z.~Perk{\'o}, S.~R. van~der Voort, S.~van~de Water, C.~M.~H. Hartman,
  M.~Hoogeman, and D.~Lathouwers.
\newblock Fast and accurate sensitivity analysis of {{IMPT}} treatment plans
  using {{Polynomial Chaos Expansion}}.
\newblock {\em Physics in Medicine and Biology}, 61(12):4646--4664, May 2016.

\bibitem{perlTOPASInnovativeProton2012}
J.~Perl, J.~Shin, J.~Schumann, B.~Faddegon, and H.~Paganetti.
\newblock {{TOPAS}}: An innovative proton {{Monte Carlo}} platform for research
  and clinical applications.
\newblock {\em Medical Physics}, 39(11):6818--6837, Nov. 2012.

\bibitem{perthameTransportEquationsBiology2007}
B.~Perthame.
\newblock {\em Transport {{Equations}} in {{Biology}}}.
\newblock Frontiers in {{Mathematics}}. {Birkh\"auser Basel}, 2007.

\bibitem{peshkinChristianSheltonLearning1992}
L.~Peshkin and C.~R. Shelton.
\newblock Learning from scarce experience.
\newblock In {\em Proceedings of the Nineteenth International Conference on
  Machine Learning}, pages 498--505, 2002.

\bibitem{pflugfelderWorstCaseOptimization2008}
D.~Pflugfelder, J.~J. Wilkens, and U.~Oelfke.
\newblock Worst case optimization: A method to account for uncertainties in the
  optimization of intensity modulated proton therapy.
\newblock {\em Physics in medicine and biology}, 53(6):1689--1700, Mar. 2008.

\bibitem{poetteGPCintrusiveMonteCarloScheme2018}
G.~Po{\"e}tte.
\newblock A {{gPC}}-intrusive {{Monte}}-{{Carlo}} scheme for the resolution of
  the uncertain linear {{Boltzmann}} equation.
\newblock {\em Journal of Computational Physics}, May 2018.

\bibitem{shirakawaSampleReuseImportance2018}
S.~Shirakawa, Y.~Akimoto, K.~Ouchi, and K.~Ohara.
\newblock Sample reuse via importance sampling in information geometric
  optimization.
\newblock {\em arXiv preprint arXiv:1805.12388}, 2018.

\bibitem{sobol1967distribution}
I.~Sobol'.
\newblock On the distribution of points in a cube and the approximate
  evaluation of integrals.
\newblock {\em USSR Computational Mathematics and Mathematical Physics},
  7(4):86--112, 1967.

\bibitem{spanierMonteCarloPrinciples1969}
J.~Spanier and E.~M. Gelbard.
\newblock {\em Monte {{Carlo}} Principles and Neutron Transport Problems}.
\newblock {Addison-Wesley Pub. Co.}, {Reading, Mass.}, 1969.

\bibitem{stammerEfficientUncertaintyQuantification2021}
P.~Stammer, L.~Burigo, O.~J{\"a}kel, M.~Frank, and N.~Wahl.
\newblock Efficient uncertainty quantification for {{Monte Carlo}} dose
  calculations using importance (re-)weighting.
\newblock {\em Physics in Medicine \& Biology}, 66(20):205003, Oct. 2021.

\bibitem{supanitskyEffectMultipleReusing2008}
A.~D. Supanitsky and G.~{Medina-Tanco}.
\newblock Effect of multiple reusing of simulated air showers in detector
  simulations.
\newblock {\em Astroparticle Physics}, 30(5):264--269, 2008.

\bibitem{tukeyConfiguralPolysampling1987}
J.~W. Tukey.
\newblock Configural polysampling.
\newblock {\em Siam Review}, 29(1):1--20, 1987.

\bibitem{unkelbachAccountingRangeUncertainties2007}
J.~Unkelbach, T.~C. Chan, and T.~Bortfeld.
\newblock Accounting for range uncertainties in the optimization of intensity
  modulated proton therapy.
\newblock {\em Physics in Medicine \& Biology}, 52(10):2755, 2007.

\bibitem{vassilievBoltzmannEquation2017}
O.~N. Vassiliev.
\newblock The {{Boltzmann Equation}}.
\newblock In O.~N. Vassiliev, editor, {\em Monte {{Carlo Methods}} for
  {{Radiation Transport}}: {{Fundamentals}} and {{Advanced Topics}}},
  Biological and {{Medical Physics}}, {{Biomedical Engineering}}, pages
  49--104. {Springer International Publishing}, {Cham}, 2017.

\bibitem{vilhenaParticularSolutionSN1995}
M.~T. Vilhena, C.~F. Segatto, and L.~B. Barichello.
\newblock A particular solution for the {{SN}} radiative transfer problems.
\newblock {\em Journal of Quantitative Spectroscopy and Radiative Transfer},
  53(4):467--469, Apr. 1995.

\bibitem{wahlEfficiencyAnalyticalSamplingbased2017}
N.~Wahl, P.~Hennig, H.-P. Wieser, and M.~Bangert.
\newblock Efficiency of analytical and sampling-based uncertainty propagation
  in intensity-modulated proton therapy.
\newblock {\em Physics in Medicine and Biology}, 62(14):5790--5807, June 2017.

\bibitem{wengVectorizedMonteCarlo2003}
X.~Weng, Y.~Yan, H.~Shu, J.~Wang, S.~Jiang, and L.~Luo.
\newblock A vectorized {{Monte Carlo}} code for radiotherapy treatment planning
  dose calculation.
\newblock {\em Physics in medicine and biology}, 48:N111--20, May 2003.

\bibitem{wieserDevelopmentOpensourceDose2017}
H.-P. Wieser, E.~Cisternas, N.~Wahl, S.~Ulrich, A.~Stadler, H.~Mescher, L.-R.
  M{\"u}ller, T.~Klinge, H.~Gabrys, L.~Burigo, A.~Mairani, S.~Ecker,
  B.~Ackermann, M.~Ellerbrock, K.~Parodi, O.~J{\"a}kel, and M.~Bangert.
\newblock Development of the open-source dose calculation and optimization
  toolkit {{matRad}}.
\newblock {\em Medical physics}, 44(6):2556--2568, June 2017.

\bibitem{wieserImpactGaussianUncertainty2020}
H.-P. Wieser, C.~P. Karger, N.~Wahl, and M.~Bangert.
\newblock Impact of {{Gaussian}} uncertainty assumptions on probabilistic
  optimization in particle therapy.
\newblock {\em Physics in Medicine \& Biology}, 65(14):145007, July 2020.

\bibitem{xiuHighOrderCollocationMethods2005}
D.~Xiu and J.~S. Hesthaven.
\newblock High-{{Order Collocation Methods}} for {{Differential Equations}}
  with {{Random Inputs}}.
\newblock {\em SIAM Journal on Scientific Computing}, 27(3):1118--1139, Jan.
  2005.

\end{thebibliography}

\end{document}